# Methane clathrates in the Solar System

Olivier Mousis

Aix Marseille Université, CNRS, LAM (Laboratoire d'Astrophysique de Marseille) UMR 7326, 13388, Marseille, France

Center for Radiophysics and Space Research, Space Sciences Building, Cornell University, Ithaca, NY 14853, USA

Université de Franche-Comté, Institut UTINAM, CNRS/INSU, UMR 6213, Observatoire des Sciences de l'Univers THETA, BP 1615, F-25010 Besançon Cedex, France

Eric Chassefière

Université Paris-Sud, Laboratoire GEOPS, CNRS, UMR 8148, Université Paris-Sud, Bât. 504, Orsay F-91405, France

Nils G. Holm

Department of Geological Sciences, Geochemistry Section, Stockholm University

Alexis Bouquet

Department of physics & Astronomy, University of Texas at San Antonio, Texas, USA

Space Science and Engineering Division, Southwest Research Institute, San Antonio, TX 78228, USA

Jack Hunter Waite

Space Science and Engineering Division, Southwest Research Institute, San Antonio, TX 78228, USA

Wolf Dietrich Geppert

Department of Physics, AlbaNova University Center, Stockholm University, Roslagstullsbacken 21, Stockholm SE-10691, Sweden

Sylvain Picaud

Université de Franche-Comté, Institut UTINAM, CNRS/INSU, UMR 6213, Observatoire des Sciences de l'Univers THETA, BP 1615, F-25010 Besançon Cedex, France

Yuri Aikawa

Department of Earth and Planetary Sciences, Kobe University, 657-8501 Kobe, Japan

Mohamad Ali-Dib
Université de Franche-Comté, Institut UTINAM, CNRS/INSU, UMR 6213, Observatoire des
Sciences de l'Univers THETA, BP 1615, F-25010 Besançon Cedex, France

Jean-Luc Charlou
Département Géosciences Marines, IFREMER-Centre de Brest, Plouzané, France

Philippe Rousselot
Université de Franche-Comté, Institut UTINAM, CNRS/INSU, UMR 6213, Observatoire des
Sciences de l'Univers THETA, BP 1615, F-25010 Besançon Cedex, France

**Corresponding Author:**

Olivier Mousis

olivier.mousis@lam.fr

Aix Marseille Université, CNRS, LAM (Laboratoire d'Astrophysique de Marseille) UMR 7326, 13388, Marseille, France

**ABSTRACT.**

We review the reservoirs of methane clathrates that may exist in the different bodies of the Solar System. Methane was formed in the interstellar medium prior to having been embedded in the protosolar nebula gas phase. This molecule was subsequently trapped in clathrates that formed from crystalline water ice during the cooling of the disk and incorporated in this form in the building blocks of comets, icy bodies, and giant planets. Methane clathrates may play an important role in the evolution of planetary atmospheres. On Earth, the production of methane in clathrates is essentially biological, and these compounds are mostly found in permafrost regions or in the sediments of continental shelves. On Mars, methane would more likely derive from hydrothermal reactions with olivine-rich material. If they do exist, martian methane clathrates would be stable only at depth in the cryosphere and sporadically release some methane into the atmosphere via mechanisms that remain to be determined. In the case of Titan, most of its methane probably originates from the protosolar nebula, where it would have been trapped in the clathrates agglomerated by the satellite's building blocks. Methane clathrates are still believed to play an important role in the present state of Titan. Their presence is invoked in the satellite's subsurface as a means of replenishing its atmosphere with methane via outgassing episodes. The internal oceans of Enceladus and Europa also provide appropriate thermodynamic conditions that allow formation of methane clathrates. In turn, these clathrates might influence the composition of these liquid reservoirs. Finally, comets and Kuiper Belt Objects might have formed from the agglomeration of clathrates and pure ices in the nebula. The methane observed in comets would then result from the destabilization of clathrate layers in the nuclei concurrent with their approach to perihelion. Thermodynamic equilibrium calculations show that methane-rich clathrate layers may exist on Pluto as well.

## 1. Introduction

Methane is ubiquitous in the universe. Its presence has been inferred in the interstellar medium (Irvine, 1987), and it is widely observed in many extraterrestrial objects such as the brown dwarfs (Oppenheimer *et al.*, 1995), the giant planets of our Solar System (Irwin 2003) and beyond (Visscher and Moses, 2011), the icy moons Titan (Owen *et al.*, 1975) and Triton (Stansberry *et al.*, 1996), the dwarf planet Pluto (Cruikshank *et al.*, 1976), and a large sample of comets (Bockelée-Morvan *et al.*, 2004). Small amounts of methane have also been observed via remote sensing in the martian atmosphere (Formisano *et al.*, 2004; Krasnopolsky, 2012; Villanueva *et al.*, 2013) and recently observed by in situ measurements onboard MSL Curiosity (Webster *et al.*, 2015). On Earth, methane is the main compound of natural gas and the second most active greenhouse gas after carbon dioxide for which it is urgent to reduce emissions.

In the Solar System, methane exists in gaseous, liquid, and pure crystalline ice forms, as a result of the wide range of thermodynamic conditions prevailing in many small bodies and planets. There is another type of crystalline ice, usually called clathrate hydrate or clathrate, which can incorporate methane and may also be widespread in the Solar System (see Kargel and Lunine, 1998 for a review). Clathrates are a class of compounds in which water forms a structure with small cages that trap guests, such as methane, that stabilize the water lattice. The two most common clathrate structures found in natural environments are known as structures I and II, which differ in the type of water cages present in the crystal lattice (Sloan and Koh, 2008). Structure I possesses two types of cages, namely, a small pentagonal dodecahedral cage, denoted $5^{12}$ (12 pentagonal faces in the cage), and a large tetrakaidecahedral cage, denoted $5^{12}6^2$ (12 pentagonal faces and 2 hexagonal faces in the cage). Structure II also possesses two types of cages, a small $5^{12}$ cage and a large hexakaidecahedral cage, denoted $5^{12}6^4$ (12 pentagonal faces and 4 hexagonal faces in the cage). The type of structure that crystallizes depends largely on the size and shape of the guest molecule. For example, methane and carbon dioxide induce water to form structure I clathrate (see Fig. 1), whereas nitrogen induces formation of structure II clathrate (Sloan and Koh, 2008).

In this paper, we review the reservoirs of methane clathrates that may exist in the different bodies of the Solar System. Methane clathrates might play an important role in the evolution of planetary atmospheres and climates. On Earth, the production of methane ultimately found in clathrates is essentially biological, and these compounds are mainly found in permafrost regions or in the sediments of continental shelves. Clathrates present in Mars and Titan's subsurfaces are suspected to continuously or sporadically feed their atmospheres in methane. If this clathrate-release of methane did not occur, these atmospheres would have been totally devoid of methane for billions of years. Clathrates may also form in the internal oceans of Enceladus and Europa and influence their compositions. More generally, locations beyond the snowline in the protosolar nebula should be the appropriate place for the formation of clathrates and their agglomeration in the building blocks of icy bodies.

In Section 2, we depict the formation processes of methane in the interstellar medium. We also describe how this molecule was subsequently embedded in the gas phase of the protosolar nebula before having been ultimately trapped in clathrate form in the building blocks of icy bodies and giant planets during disk evolution. Section 3 is devoted to a discussion of the formation, storage, and destabilization conditions of methane clathrates on Earth and Mars, and their possible impact on climate evolution. In Section 4, we review the current scenarios of Titan's formation and evolution, as they all support the idea that methane clathrates played, and continue to play, an important role in the shaping of the satellite's current properties. We also address the cases of Enceladus and Europa's internal oceans where clathrates might form. We finally discuss the roles of methane-rich clathrates in the formation and evolution of comets, Pluto/Triton, and Kuiper Belt Objects. Section 5 is devoted to conclusions.

## 2. The fate of methane from interstellar medium to the protosolar nebula

### 2.1 Methane formation in the interstellar medium

A rich chemistry of hydrocarbons is observed in the interstellar medium (Irvine, 1987; Herbst and Klemperer, 1973; Prasad and Huntress, 1980), and methane is one of the fundamental hydrocarbon molecules in this environment. The observation of gaseous methane, however, is not easy; the molecule lacks a permanent dipole moment and thus cannot be detected by radio observations of pure rotational spectral lines – a method through which many other interstellar molecules have been identified. Alternatively, gaseous methane has been detected via absorption when viewing towards high mass young stellar objects (Lacy *et al*., 1991; Knez *et al*., 2009), with a $CH_4$ abundance of order $10^{-7}$-$10^{-6}$ relative to $H_2$. Unfortunately, such observations are, so far, possible only towards very bright sources. In low-mass (Sun-like) star forming regions, Sakai *et al*. (2009) detected $CH_3D$. Assuming a typical molecular D/H ratio in low-mass star-forming regions, the gaseous $CH_4$ abundance is derived to be $10^{-7}$.

For several molecules in dark clouds, including $CH_4$ and $H_2O$, a pathway starting with successive hydrogenation of ions by $H_2$ and concluding with the dissociative recombination of the protonated form of the respective neutral has been invoked. In the case of methane, this sequence would start with the reaction of carbon atoms

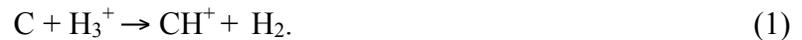

$$C + H_3^+ \rightarrow CH^+ + H_2. \qquad (1)$$

Successive hydrogenation then leads to the formation of $CH_3^+$ ions:

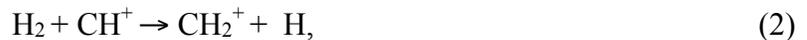

$$H_2 + CH^+ \rightarrow CH_2^+ + H, \qquad (2)$$

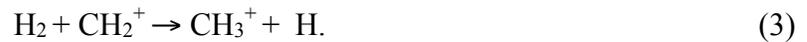

$$H_2 + CH_2^+ \rightarrow CH_3^+ + H. \qquad (3)$$

The product of Reaction 3 (the $CH_3^+$ ion) is relatively stable and cannot abstract hydrogen from $H_2$. Thus a radiative recombination process has to be invoked to form $CH_5^+$ ions

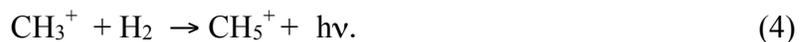

$$CH_3^+ + H_2 \rightarrow CH_5^+ + h\nu. \qquad (4)$$

The rate constant of this reaction is $3.78 \times 10^{-16}$ $(T/298)^{-2.3}$ exp(-21.5/T) cm$^3$ s$^{-1}$, which is $1.1 \times 10^{-13}$ cm$^3$ s$^{-1}$ at 10K (Talbi and Bacchus-Montabonel, 1998; KIDA database[1]).

$CH_5^+$ can then undergo proton transfer to CO to form methane:

$$CH_5^+ + CO \rightarrow CH_4 + HCO^+. \tag{5}$$

If CO is not abundant, dissociative recombination can function as the final step:

$$CH_5^+ + e^- \rightarrow CH_4 + H. \tag{6}$$

But reaction (6) is only a minor pathway in the dissociative recombination of $CH_5^+$ (branching fraction 0.05) (Semaniak *et al.*, 1998; Kaminska *et al.*, 2010). In spite of the apparent inefficiency of reactions (4) and (6), time-dependent models of gas-phase chemistry in a molecular cloud can make $CH_4$ as abundant as $10^{-6}$ (relative to $H_2$) at a few $10^5$ yr and $10^{-8}$ at $10^8$ yr (Millar *et al.*, 1997; Woodall *et al.*, 2007; Aikawa, 2013).

Sequential hydrogenation of carbon atom on cold grain surface is an alternative and probably more efficient mechanism to form methane:

$$C + 4H \rightarrow CH_4 \tag{7}$$

Formation of methane by impinging H atoms on graphite has been experimentally proven to occur at very low temperatures ($T \geq 7$ K) (Bar-Nun *et al.*, 1980). Observation of methane ice via absorption when observing in the direction of high-mass young stellar objects indicates $CH_4$ ice abundance is a few percent relative to $H_2O$ ice (Lacy *et al.*, 1991; Boogert *et al.*, 1996). Assuming a canonical abundance of $\sim 10^{-4}$ for $H_2O$ ice, $CH_4$ ice abundance is $10^{-6}$ relative to $H_2$. The Spitzer space telescope detected $CH_4$ ice features in the direction of 25 out of 52 low-mass young stellar objects. The $CH_4$ ice abundance from these observations ranged from 2 to 13 percent relative to $H_2O$ ice (Oberg *et al.*, 2008). In gas-grain models of low-mass dense cores, $CH_4$ ice can indeed be as abundant as a few to 10 percent relative to water ice (Aikawa *et al.*, 2005). When a forming protostar heats the molecular cloud core, a fraction of $CH_4$ ice is sublimated, which reacts with $C^+$ to restart rich hydrocarbon chemistry in the gas phase (Sakai *et al.*, 2008, 2009; Aikawa *et al.*, 2008; 2012).

## 2.2    Methane incorporation in clathrates formed in the protosolar nebula

Enclathration is a mechanism that would favor the efficient trapping of $CH_4$ and other volatiles in planetesimals. Indeed, formation scenarios of the protosolar nebula invoke two main reservoirs of ices that took part in the production of icy planetesimals (see Fig. 2). The first reservoir, located within 30 AU of the Sun, contains ices (dominated by $H_2O$, $CO$, $CO_2$, $CH_4$, $N_2$, and $NH_3$) that originated from the interstellar medium, which, due to their near solar vicinity, were initially vaporized. Because the kinetics is inhibited, chemical models predict that the gas phase ratios $CO/CH_4$ and $N_2/NH_3$ in the protosolar nebula remained almost similar to those initially acquired in the interstellar medium (Lewis and Prinn, 1980; Prinn and Fegley, 1989). In these conditions, $CO$ and $N_2$ remained the two main C- and N-bearing volatiles in the protosolar nebula gas phase, with $CH_4$ and $NH_3$ one order of magnitude less abundant (Mousis *et al.*, 2002a). These results, which still remain valid, are at odds with previous studies predicting that $CH_4$ and $NH_3$ were the dominant species from equilibrium models of the protosolar nebula (Grossman, 1972; Lewis, 1972). With time, the decrease of temperature and pressure conditions allowed the water in this reservoir to condense at ~150 K in the form of microscopic crystalline ice (Kouchi *et al.*, 1994) at typical pressures met in the midplane of the disk. This location corresponds to the position of the so-called snowline (or water iceline) in the protosolar nebula (Cyr *et al.*, 1998; Lunine 2006). It is postulated that a substantial fraction of the volatile species were trapped as clathrates during this condensation phase as long as free water ice was available within 30 AU in the outer solar nebula (Mousis *et al.*, 2000). On the other hand, the remaining volatiles that were not enclathrated (due to the lack of available water ice) probably formed pure condensates at lower temperatures in this part of the nebula (Mousis *et al.*, 2012a,b). The formation locations of methane clathrates or pure condensates would also correspond to the positions of methane icelines in the protosolar nebula. The other reservoir, located at larger heliocentric distances, is composed of ices that originated from the interstellar medium and did not vaporize when entering into the disk. In this reservoir, water ice was essentially in the amorphous form, and the other volatiles remained trapped in the amorphous matrix (Owen *et al.*, 1999; Notesco and Bar-Nun, 2005). Consequently, icy solids that formed at heliocentric distances less than 30 AU mainly agglomerated from a mixture of clathrates and pure condensates, whose ratio depends on the amount of available crystalline water and its clathration efficiency (Mousis *et al.*, 2009b). In contrast, solids produced at higher heliocentric distances (i.e., in the cold outer part of the solar nebula) were formed from primordial amorphous ice originating from the interstellar medium. Thus, clathrates may have been agglomerated in comets (Marboeuf *et al*., 2010, 2011, 2012), in the building blocks of the giant planets (Gautier *et al*., 2001; Alibert *et al*., 2005a,b; Mousis *et al*., 2009b, 2012a), and in their surrounding satellite systems (Lunine and Stevenson, 1985; Mousis, 2004; Mousis and Gautier, 2004; Mousis *et al*., 2009a). Even if the global picture regarding the possible existing ice forms in the protosolar nebula is well understood, one must note that there is an important deficiency of clathrate equilibrium data at pressures well below 1 bar (Sloan and Koh, 2008; Fray *et al*., 2010). Typical nebular pressures are in the $10^{-9}$-$10^{-6}$ bar range, and the partial pressures of concerned volatiles is at minimum 3 or 4 orders of magnitude lower. This implies the extrapolation of laboratory equilibrium data obtained typically in the mbar range down to very low pressure to derive the formation conditions of clathrates in the protosolar nebula.

### 2.2.1 The water abundance throughout the protosolar nebula

Before going further, it is necessary to discuss the plausible abundance profile of crystalline water, which might have existed during this condensation epoch throughout the nebula and is needed for clathrate formation. Interestingly, although it remains challenging to derive the spatial distribution of water ice abundance in the midplanes of protoplanetary disks from observations (Thi *et al*., 2002; Pontoppidan *et al*., 2005; Terada *et al*., 2007; Aikawa *et al*., 2012), gaseous water lines have been detected in their upper layers (Carr and Najita, 2011; Hogerheijde *et al*., 2011; Zhang *et al*., 2013).

At low pressure, water vapor condenses beyond the so-called "snowline," which corresponds to the region where the temperature drops below 150 K. Inside the snowline, water is in vapor form, but it exists only as ice beyond the snowline. Several studies suggest that the abundance of water is not homogeneous throughout the protosolar nebula. Stevenson and Lunine (1988) proposed that the outward vapor diffusion induced by the presence of the snowline with its subsequent condensation in a narrow location (with a width around 0.5 AU) would deplete the region inside the snowline of vapor down to subsolar values, and increase substantially the ice abundance at the snowline location by factors 5 to 10 times the protosolar value (assuming that all O forms

H$_2$O in the protosolar nebula). More recently, Ali-Dib *et al*. (2014) calculated that this depletion would also take place with modern realistic replenishment rates (due to the inward drift and sublimation of the particles created from the condensation of gas at the snowline) and assuming that solids in the protosolar nebula are almost entirely in the form of decimetric pebbles, a size thought to be optimal for accretion timescales. In this model, the vapor diffusing from the inner nebula condenses at the snowline onto existing submillimetric dust that starts growing through coagulation and condensation. Once the pebble size is reached, the dust migrates inward and sublimates inside the snowline, and therefore replenishes the depleted vapor. Because the replenishment timescale is longer than diffusion timescale, this replenishment is not fast enough to counter the diffusion, which leads to a depleted inner nebula in $\sim 10^4$ yr. On the other hand, the ice abundance beyond the enriched region should remain more or less in protosolar abundance with respect to H$_2$ gas because water ice is homogeneously distributed beyond the snowline, except in the neighborhood of planetary feeding zones. Figure 3 shows the water abundance profile throughout the protosolar nebula derived from the model of Ali-Dib *et al*. (2014). It shows that the inner nebula is almost completely depleted in water, but with a highly enriched snowline, and with an intact abundance beyond it (Ciesla and Cuzzi 2006).

### 2.2.2   *Formation of methane-rich clathrates in the outer protosolar nebula*

Clathrates form during the cooling of the outer protosolar nebula after ice crystallization and usually at higher temperatures than pure condensates (Mousis *et al*., 2009a,b; Mousis *et al*., 2010). The clathration process stops when no more crystalline water ice is available to trap the volatile species. Beyond this point only pure condensates form at a lower temperature if the disk continues to cool down. Solids formed under these conditions are then agglomerated from a mixture of pure ices, stoichiometric hydrates (such as NH$_3$-H$_2$O hydrate), and multiple guest clathrates that crystallized in the form of microscopic grains at various temperatures in the outer part of the disk. While the kinetics (i.e., formation rate) of clathrate formation is not yet well constrained in the low pressure range, the process of volatiles being trapped in icy grains may be theoretically described by using the equilibrium curves of pure condensates and of the stoichiometric hydrate NH$_3$-H$_2$O, a statistical-thermodynamic model that determines the equilibrium curves and compositions of multiple guest clathrates, and the thermodynamic paths detailing the evolution of temperature and pressure between 5 and 30 AU in the protosolar nebula (Mousis *et al*., 2012b). As mentioned above, the abundance of water is crucial for enabling clathrate formation in the protosolar nebula and for determining the nature of entrapped volatiles.

Figure 4 represents the composition of icy planetesimals as a function of their formation temperature in the outer protosolar nebula, which is computed for two different cases of water abundances. In the first case, the amount of water is derived from a protosolar oxygen abundance. The resulting water abundance is not sufficient to allow for the clathration of all volatiles present in the gas phase of the protosolar nebula, which usually occurs in the ~45-80 K temperature range at nebular pressure conditions. Therefore, the protosolar nebula has to cool down to extremely low temperatures (~20-25 K) to permit the condensation of essentially all $N_2$ and a significant fraction of CO in the form of pure ices. In the second case, the abundance of water is set high enough to allow for the trapping of all species in multiple guest clathrates, a mechanism that could have been effective in Jupiter's feeding zone (see previous Section). Here, all volatiles are captured in solids in the ~45-80 K range, including CO and $N_2$. Interestingly, in both situations, $CH_4$ is entrapped in clathrates at ~54 K, irrespective of the considered water abundance. Since the water abundance profile is not depleted beyond the snowline (see Fig. 3), methane should always be trapped in clathrate structures and never form a pure condensate at lower temperature and pressure conditions in the protosolar nebula. Consequently, methane has probably been agglomerated in clathrate form in the building blocks of comets, giant planets, and their surrounding icy satellites (Mousis *et al*., 2009a,b, 2012a).

## 3.    Methane clathrates in terrestrial planets

### 3.1.  Methane clathrates on Earth

In general, the formation of the common clathrates on Earth needs four ingredients: high pressure, low temperature, an abundance of easily degradable organic matter, and plenty of water. An additional requirement is that the fluid phase needs to be saturated with respect to methane. If the ambient temperature is low, as it is in the Arctic and Antarctic, a pressure close to surface conditions may be sufficient to stabilize the clathrate. This enables the thermodynamic stability of methane clathrates in terrestrial permafrost regions. The production of methane that leads to formation of such clathrates is dominantly biological and depends primarily on fermentation of organic material and anaerobic oxidation of organic material or molecular hydrogen with carbon dioxide as the terminal electron acceptor, in which the carbon is reduced to form methane. In these biological processes – the anaerobic oxidation in particular – the microorganisms have a pronounced preference for the light carbon isotope $^{12}C$. The methane that is formed may, therefore, get extremely light isotopic signatures down to $\delta^{13}C = -70‰$. Because of the requirement of organic material or hydrogen the prevailing regions of oceanic methane clathrates on Earth occur in the sediments on the continental shelves at some depth beneath the sea surface in high-productivity oceanic upwelling areas (see Fig. 5). The specific depth depends on the ambient temperature. Methane clathrates, once formed, are thermodynamically stable also in sediments of the abyssal plains of the deep-sea. No clathrates are, however, formed there because of the extremely low production rate of organic matter in the abyssal ocean.

We normally think of clathrates as being mostly methane gas with some additional water. This is partly true: destabilization of 1 $m^3$ of methane clathrate at standard temperature and pressure yields 164 $m^3$ of gas and 0.8 $m^3$ of water (Kvenvolden, 1998). However, if we think of it the other way around, out of 1 $m^3$ of solid clathrate 0.8 $m^3$ is water, while only 0.2 $m^3$ is methane that has been locked in the structure of this solid compound. Thus, the requirement of water during the formation of methane clathrate is immense. The availability of water may be a problem in finely grained and tight clayey sediment. What often happens is that once a seed of clathrate forms, it very slowly expands and pushes away the sediment. If several such seeds grow large enough they may merge and form a continuous horizon of clathrates within the sediment.

The global inventory of methane clathrate in marine sediments is not precisely known. It has been estimated from models representing the seafloor area according to depth, temperature, and $O_2$ concentration, and making assumptions on the sedimentation rate of organic carbon as a function of water depth, and its burial rate in sediments. In this way, steady state values of methane in its different states—dissolved in sea water, frozen in seafloor clathrates, gaseous in methane bubbles—may be calculated. The resulting calculation indicates ~3000 Gt of carbon in clathrates and ~2000 Gt of carbon in bubbles (Buffet and Archer, 2004). Higher values of the frozen methane inventory have been reported in the literature, up to 20,000 Gt of carbon (see e.g. Higgins and Schrag, 2006, and references therein). This range may be compared to the entire world's conventional oil and gas inventory of ~5,000 Gt of carbon making sea floor methane a possible resource for the future.

An increase of temperature can destabilize methane clathrates and release methane. The free gas that is released and accumulated underneath marine clathrates may break through the solid layer because of earthquakes and other physical disturbances. Such events may cause huge submarine landslides that would release even more $CH_4$ to the ocean water and to the atmosphere (see Fig. 5). This methane may have climatic effects as a greenhouse gas, while the landslides may cause major tsunamis with severe damage of coastal areas around the world's oceans. For example, the Paleocene-Eocene Thermal Maximum, which occurred ~55 Ma, has been attributed to a sudden release of greenhouse gases (carbon dioxide and/or methane) to the ocean/atmosphere reservoir (see e.g. Dickens *et al.*, 1995). During more than 10,000 years, Earth's surface temperature increased by 5-10°C, and bottom water temperature by more than 4°C. A number of isotope records, based on data of primary carbonates and organic matter, show a dramatic drop in the ratio of $^{13}C$ to $^{12}C$, with a $\delta^{13}C$ value of 2-to-3‰, in the global ocean/atmosphere inorganic carbon reservoir at this epoch. This shift would require the addition of at least 30,000 Gt of mantle $CO_2$ with a $\delta^{13}C$ value of -7‰. Such a high value, similar to the present $CO_2$ ocean/atmosphere reservoir, is highly implausible. Due to the small $\delta^{13}C$ value in methane clathrates of -60‰, Paleocene-Eocene Thermal Maximum alternatively may have been rather triggered by the so-called *clathrate gun hypothesis* (Kenneth *et al.*, 2000, 2003; Renssen *et al*., 2004), which corresponds to an episode of release of a few 1,000 Gt of carbon originating in methane clathrates. The relatively low values of ~1,000 to 2,000 Gt of carbon first calculated by Dickens *et al.* (1995) have been revised upward in some later works (e.g., ~3,000 to 6,000 Gt of carbon, Sexton *et al.*, 2011). Today's global inventory of methane in seafloor clathrates should have been one half that of the Paleocene because of a warmer climate and therefore does not exceed 2000 Gt of carbon (Buffet and Archer, 2004). If so, the global inventory of carbon in methane clathrates at the Paleocene seems too low to explain the Paleocene-Eocene Thermal Maximum. Today, the clathrate gun hypothesis is no longer viewed credibly in its most extreme form (Sowers, 2006; Severinghaus *et al.*, 2006), but this mechanism is recognized as a likely contributor to the immense climate warming of the Earth's natural past. Additionally, the triggering mechanism of physical clathrate destabilization is not identified. Alternatively, an episode of submarine volcanism may have resulted in the intrusion of huge mantle-derived melts in carbon-rich sedimentary layers in the northeast Atlantic, with explosive release of methane to the ocean/atmosphere reservoir (Svensen *et al.*, 2004).

Although clathrate destabilization is not firmly established as the cause of the Paleocene-Eocene Thermal Maximum, this example shows the potentially important role of methane clathrates in climate evolution, as could be the case on Mars (see 3.2 Section below). On present Earth, the presence of large amounts of clathrates is primarily a threat with respect to terrestrial and shallow marine clathrates in permafrost. Deep marine clathrates are much less sensitive to changes in temperature because of the high hydrostatic pressure (Kvelvolden, 1998). Most existing deep marine clathrates on Earth are not in the vicinity of the border of the pressure-temperature stability field for clathrates. A sea-level rise, for instance, caused by global warming and melting of grounded polar ice caps would increase the stability of the clathrates. A potential sea-level regression, on the other hand, would primarily destabilize the clathrate at its base because of interaction with the geothermal gradient, making the layer of clathrate thinner. Interestingly, it should be noted that the industrial exploitation of methane clathrates in sea floors should be less harmful for the environment than shale gas. Indeed, accidental leakage of shale gas alone is a significant part of recent anthropogenic warming. Because of that, the footprint for shale gas is greater than that for conventional gas or oil on any time horizon. Compared to coal, the footprint of shale gas will likely be more than twice as great over the course of the next twenty years (Howarth *et al.*, 2011).

### 3.2. Methane clathrates on Mars

Mars is a cold planet. The seasonally averaged temperature at the surface is ~220 K and the average atmospheric pressure is ~7 mbar. The composition of the atmosphere is dominated by $CO_2$ (96%), $N_2$ (2%), Ar (2%) with traces of other compounds: $O_2$, CO, $H_2O$, $H_2$, $O_3$, and $CH_4$ (at low level: a few ppb to a few tens ppb) (Owen *et al.*, 1977; Mumma *et al.*, 2009; Mahaffy *et al.*, 2013). The martian climate is very contrasted. Its surface pressure covers the 3-10 mbar range, depending on seasonal conditions and altitude, and its surface temperature may reach values as low as 140 K in winter on the south polar cap. A water ice cryosphere extends typically to 10 km in depth (in the ~0 to 20 km range), depending on latitude and crust physical properties, below the surface (Lasue *et al*., 2013, and references therein). This cryosphere is potentially rich in clathrates of various gases, either originating from the atmosphere or produced at depth by magmatic/hydrothermal processes and migrating up to deep cryospheric layers. The discovery of $CH_4$ in the atmosphere of Mars by the PFS instrument on the Mars-Express orbiter (Formisano *et al*., 2004) has drawn attention to the possible role of clathrates as a large scale cryospheric reservoir of this gas, which could have been produced in large amounts in the past and remained trapped, at least partially, in the crust or inside polar caps until recent epochs. Such reservoirs present in the close subsurface would allow the sporadic release of $CH_4$ into the atmosphere through dissociation or ablation (Kargel, 2004; Prieto-Ballesteros *et al*., 2006; Chastain and Chevrier, 2007; Mousis *et al*., 2013a), explaining its presence in the atmosphere. Methane clathrate has also been considered to be a likely source of thermal insulation in parts of the crust, and it has been proposed that such insulators could drive further geologic activity (Kargel *et al*., 2007). Earlier, Ross and Kargel (1998) reviewed the thermal conductivity of clathrate hydrates and proposed some applications to the possible thermal state of martian polar caps. In addition, Kargel and Lunine (1998) reviewed what was then known of clathrates in the Solar System (including methane clathrates) and considered the possible stability and occurrence of clathrates on Mars, although the focus was on $CO_2$ clathrate in that work. Finally, Max and Clifford (2000), Kargel *et al*. (2000), Pellenbarg *et al*. (2003), and Kargel (2004) discussed some very wide-ranging geological and biological roles of methane clathrates on Mars, if they occur.

### 3.2.1   *Observation of $CH_4$ in the Martian atmosphere*

$CH_4$ has been detected in Mars' atmosphere at the 10-60 ppb level by several instruments from space and ground-based observatories at the end of the nineties and during the following decade, spanning ~6 Martian years (MY24-MY29) (Krasnopolsky *et al*., 2004; Formisano *et al*., 2004; Mumma *et al*., 2009; Fonti and Marzo, 2010). Recent published $CH_4$ observations (Krasnopolsky, 2012; Villanueva *et al*., 2013) suggest a small $CH_4$ mixing ratio of ~10 ppb, and in some cases no or little $CH_4$ with an upper limit of ~7 ppb in 2009-2010, during Mars' norther spring. More recent MSL Curiosity in situ measurements are showing variations in the methane detection at Gale Crater over the 8 km traverse to Mount Sharp during the two terrestrial years of the nominal MSL mission. While the background level of methane remains around 0.69 ± 0.25 ppb, an elevated level of methane of 7.2 ± 2.1 ppb was detected during approximately 2 months (Webster et al. 2015), comparable to typical levels observed by remote sensing at the end of the last decade. All measurements suggest that $CH_4$ is highly variable both in space and time. Based on the existing photochemical scheme of atmospheric $CH_4$ oxidation, such a variability would require a photochemical lifetime of less than 1 Martian year (Lefèvre and Forget, 2009), much shorter than the conventional lifetime of 300 yr based on existing photochemical models. Such a short lifetime has been considered by some as implausible. Therefore, it has been alternatively argued that Mars' $CH_4$ lines observed from Earth might be strongly affected by competing telluric $CH_4$ lines, bringing into question the reality of the detection from ground-based observatories (Zahnle *et al.*, 2011). Recent MSL in situ measurements confirm both presence of methane and typical mixing ratios of ≈10 ppb during active periods, together with abrupt time and space variations. To explain a short $CH_4$ lifetime, it has been suggested that $CH_4$ molecules in contact with the regolith could be adsorbed at the surface of ferric mineral grains whose superficial layers are permanently oxidized by atmospheric $H_2O_2$ and other oxidants, making surface oxidation a possible sink for $CH_4$ (Chassefière and Leblanc, 2011a). Further experimental work is required to assess the plausibility of this hypothesis. Note that direct oxidation of $CH_4$ by $H_2O_2$ at the surface of grains has been proved to be inefficient at low Mars temperatures (Gough *et al.*, 2011).

### 3.2.2. Possible origins for $CH_4$ on Mars

The origin of $CH_4$ is still largely unknown. An external meteoritic source, as well as a direct magmatic origin, do not seem capable of explaining the observed amount of $CH_4$ in the atmosphere (Atreya *et al.*, 2007). $CH_4$ may rather be produced by deep geochemical processes such as hydrothermal alteration of basaltic crust (Lyons *et al.*, 2005) or by serpentinization of olivine-rich material producing $H_2$ and reducing crustal carbon into $CH_4$ through the reaction of dissolved $CO_2$ with $H_2$. This process has been observed in hydrothermal systems at mid-ocean ridges (Oze and Sharma, 2005). Possibly, if life developed at depth on Mars, some of the released $CH_4$ could be biogenic, as observed on mid-ocean ridges and in geothermal systems on Earth (Emmanuel and Ague, 2007). By studying terrestrial ophiolitic rocks, which are able to generate large amounts of $CH_4$ during serpentinization, it has been recently suggested that serpentinized    rocks on Mars could produce significant amounts of $CH_4$ at relatively low temperatures  (<50°C) (Etiope *et al*., 2013). Such a mechanism could explain the production of $CH_4$ on Mars in the ancient past or in the present (see Holm *et al*., 2014 for details).

It is difficult to estimate the amount of $CH_4$ that could have been formed via this process during Mars' history. The amount of water involved in serpentinization during Mars' history has been modeled and estimated to be potentially large, a several hundred meters thick water Global Equivalent Layer (Chassefière and Leblanc, 2011b) based on the constraint provided by the present atmospheric D/H ratio (~5 SMOW, Owen *et al*., 1988; Webster *et al*., 2013), which is directly related to the accumulated amount of hydrogen released in the serpentinization process to the atmosphere, then lost to space,. Because serpentinization results in the formation of ferric iron, the magnetic field measured by MGS in the Southern hemisphere (Langlais *et al*., 2004) may also be used as a constraint. The removal of a ~500 m (300-1000 m) thick water global equivalent layer by serpentinization during Mars' history fits both present atmospheric D/H and southern crust magnetization (Chassefière *et al*., 2013b). The corresponding accumulated amount of $CH_4$ released by serpentinization could be up to 3.6 x $10^{21}$ g (equivalent to a ~0.4 bar $CH_4$ partial pressure) if a 1000 m thick global equivalent layer has been consumed and all the produced hydrogen is in the form of $CH_4$. Assuming a 300 m thick global equivalent layer, and a $CH_4$ molar fraction ($CH_4/(H_2+CH_4)$) of 0.1, the accumulated mass of the produced $CH_4$ is 3.6 x $10^{20}$ g (~0.04 bar $CH_4$ partial pressure assuming a global pure-$CH_4$ uniform atmosphere).

These values, although they must be considered as an upper limit, are quite significant. Because large amounts of $CH_4$ could have been produced in subsurface hydrothermal systems during Mars history, the fate of this $CH_4$ is important. Has it been released directly to the atmosphere, with subsequent loss of hydrogen to space, or temporarily stored in the cryosphere inside clathrates? If stored in the cryosphere, what are the stability zones of $CH_4$ clathrates in the subsurface, and how long has $CH_4$ has been stored after being produced? What are the mechanisms of $CH_4$ release into the atmosphere? Although all these questions are still poorly answered, some preliminary elements of answers may be provided.

### 3.2.3   *Stability of methane clathrates and transport/storage/release processes*

The phase diagram of pure $CH_4$ clathrates is shown in Fig. 6. At the mean atmospheric conditions that prevail at the surface of Mars (*T* ~220 K, *P*~7 mbar), pure $CH_4$ clathrates are not stable and cannot form by condensation of atmospheric $CH_4$. The same is true for the main atmospheric constituent, $CO_2$, and all minor ones ($N_2$, Ar, CO, $O_2$, etc.). During winter, on the south permanent $CO_2$ ice polar cap, the temperature can be as low as 140 K. Even at such a low temperature, the equilibrium pressure is larger than 10 mbar (Fig. 6), and pure $CH_4$ clathrates are not stable. Because $CO_2$-dominated clathrates are stable below 150 K, $CH_4$ can nevertheless condense together with $CO_2$ into $CO_2$-$CH_4$ multiple guest clathrates in winter on the south polar cap, with relative abundance of $CH_4$ in the clathrate one third to one fourth of its value in the atmosphere (Thomas *et al*., 2009; Herri and Chassefière, 2012).

In addition, Fig. 6 shows that $CH_4$ clathrates are stable in the martian crust, from a few meters below the surface down to a ~10 km depth, in agreement with the conclusions found by Kargel (2004) and Prieto-Ballesteros *et al*. (2006). If, at some point early in its evolution, the atmosphere of Mars had been rich in $CH_4$ (see e.g. Kasting, 1997), and the background pressure has been of the order of several hundred millibar or more of $CO_2$, large amounts of $CH_4$ may have been trapped in shallow subsurface clathrates (Prieto-Ballesteros *et al*., 2006; Thomas *et al*., 2009). If $CH_4$ has been formed at depth in hydrothermal systems, either hydrothermally or biologically, it has been transported upward by convecting hydrothermal fluid or in gas phase, depending on the nature of the transition between the cryosphere and deep aquifers, via either a direct contact or an unsaturated zone where gas may freely circulate. Two possible paths of an ascending parcel of a $CH_4$-rich fluid are represented in the ($P$, $T$) plane on Fig. 6. At great depth, $CH_4$ is dissolved in the fluid. Then bubbles form in the ascending fluid due to pressure decrease. When reaching their stability boundaries, at 10 km and 3.5 km depths in the two considered cases, clathrates form at the interface between the gas inside the bubbles and ambient water under the form of a thin film surrounding the bubble, as observed in oceanic $CH_4$ plumes forming above deep-sea $CH_4$ sources (Sauter *et al*., 2006). When crossing the $H_2O$ solid/liquid boundary, at 7 km and 2.7 km depths in the two considered cases, clathrates solidify and fill the pore space available in the preexisting water ice, so there would not be anymore convection (Elwood Madden *et al*., 2009). Clathrates can further expand upward over time through diffusion. The removal timescale by diffusion of clathrates from a 10 km deep cryosphere is very poorly known, and could be anywhere from 2 x $10^4$ to $10^{12}$ yr (Mousis *et al*., 2013a). $CH_4$ can therefore be stored in subsurface clathrates for a long time before being further released to the atmosphere.

The storage capacity of a $CH_4$-saturated cryosphere on early Mars is quite large. According to recent calculations by Lasue *et al.* (2014) from a physical model of Mars' cryosphere, the global storage capacity of the Noachian cryosphere is $1 \times 10^{20}$ to $6 \times 10^{20}$ moles of $CH_4$ assuming a $CH_4$-clathrate saturated cryosphere. By restricting this capacity to the fraction of southern terrains where a significant remanent magnetic field is observed (<3.6 x $10^{20}$ g, see above and Chassefière *et al.*, 2013b), the methane trapping capacity of the cryosphere is $2 \times 10^{19}$ to $1.5 \times 10^{20}$ g. A quantity of serpentinization-derived methane of $2 \times 10^{19}$ g (or 20,000 Gt) or more, corresponding to an amount of water stored in serpentine of only a few tens of meters deep global equivalent laye of water, could have been stored in the cryosphere of the cold early Mars. Any sudden destabilization of the early Mars cryosphere, for example, due to impacts or obliquity variations, could have released to the atmosphere large amounts of $CH_4$, resulting in a global heating of the planet through greenhouse effects (Kasting, 1997). According to calculations by Lasue *et al.* (2014), $CH_4$ emission fluxes could have been comparable to, or larger than, those at the origin of the Paleocene-Eocene Thermal Maximum which occurred on Earth ~55 Ma (See Section 3.1).

Mechanisms of $CH_4$ release from the upper cryosphere into the atmosphere are poorly understood. The space and time variability of $CH_4$ concentration in the atmosphere is puzzling, as already mentioned. Some $CH_4$ plumes have been observed in equatorial regions, where no water ice is observed at the surface. At high latitudes, where water ice is present close to the surface, no release of $CH_4$ has been observed except over the north polar cap during northern summer (Geminale *et al.*, 2011). If $CH_4$ is mainly released from equatorial regions, where ice is below the depth of a few meters and is unaffected by seasonal temperature changes, the release flux is not controlled by seasonal cycles, which seems to contradict the fact that a seasonal trend is observed, with more $CH_4$ during northern summer and lower levels during northern winter (see Fig. 2 in Mousis *et al.*, 2013a). The space and time variability of the $CH_4$ release rate, which seems to be required to explain the observed atmospheric $CH_4$ variations, is difficult to explain. The geothermal gradient, which controls the stability of clathrates, is not expected to vary over short time scales. Sporadic release could be due to seismic or tectonic activity, and geographical variability to possible heterogeneities of the superficial cryosphere and its content in $CH_4$ clathrates. But these mechanisms cannot explain the observed global seasonal pattern. A possible obstruction by seasonally variable glacial deposits of the porous regolith, preventing $CH_4$ from being released to the atmosphere, doesn't seem compatible with identified source regions (equatorial) and the observed seasonal variation pattern. The possible condensation of $CH_4$ in $CO_2$-$CH_4$ clathrates on the south polar cap during winter (Herri and Chassefière, 2012) and the adsorption of $CH_4$ in the regolith (Meslin *et al*., 2011) are not efficient enough to generate significant seasonal variations in the atmosphere.

It has been proposed that, due to the metastability of clathrates (named "anomalous preservation," (see e.g., Stern *et al*., 2001), $CH_4$ could be released, at least partially, to the atmosphere under the form of small metastable submicron particles, which are lifted up to the middle atmosphere by convection, then transported by winds during weeks or months, before progressively decomposing to gas phase under the effect of cloud processes (Chassefière, 2009). If so, $CH_4$ plumes would form inside the atmosphere far from the release location at the surface in a sporadic way depending on cloud activity, and thus would be related to the water content of the atmosphere. To our knowledge, this hypothesis is the only one that provides a self-consistent frame for explaining the observed apparent seasonal variability, but it would require extensive laboratory experiments to be validated.

### 3.2.4   Conclusions

Other clathrates apart from $CH_4$ clathrates could have played an important role in driving Mars atmospheric composition and climate. The late deposition of sulfate minerals (Bibring *et al*., 2006) has been suggested to be due to the sequestration of volcanic $SO_2$ in the Noachian upper cryosphere under the form of planetary-scale $CO_2$-$SO_2$ clathrate reservoirs (Chassefière *et al*., 2013a). The two orders of magnitude drop between the measured atmospheric abundances of non radiogenic argon, krypton, and xenon in Earth versus Mars could be explained by the early sequestration in the martian cryosphere of these noble gases under the form of clathrates (Mousis *et al*., 2012c).

No direct observation of clathrates has been made on Mars to date. Observations planned for the Trace Gas Orbiter (TGO) of the Exomars mission, starting in 2016, should bring a definitive answer to the question of the reality of $CH_4$ in Mars' atmosphere, and if $CH_4$ is present should considerably improve our understanding of $CH_4$ variability and cycles. If metastable particles of clathrates are present in Mars' atmosphere, they could be observed by a combination of IR and UV spectroscopy from orbit, but such detection is quite challenging due to the small size of the particles and the required high spectral resolution and signal-to-noise ratio. Clathrates may be spectroscopically searched for on the icy surface of the polar caps, where $CO_2$-minor species multiple guest clathrates may form during winter. Subsurface clathrates can be directly reached only from landers by drilling and compositional analysis. Laboratory experiments are also needed to simulate the thermodynamic and kinetic behaviors of clathrates in martian conditions (Mousis *et al*., 2013a).

## 4. Methane-rich clathrates in solid bodies of the outer Solar System

As discussed in Section 2, methane is believed to have been encaged in multiple guest clathrates during the cooling of the outer part of the protosolar nebula within the inner ~30 AU. Multiple guest clathrates would have then been agglomerated with crystalline ices and taken part in the composition of the building blocks of the outer icy satellites, including Titan and Enceladus, as well as of comets or Kuiper Belt Objects.

### 4.1 Titan

Titan's methane probably originates from the protosolar nebula (Mousis *et al*., 2002b; Mousis *et al.*, 2009a). It has been proposed that, similarly to Mars, methane was synthesized via serpentinization reactions within the interior of the satellite (Atreya *et al.*, 2006), but this mechanism is invalidated by the atmospheric measurement of the methane D/H ratio (Mousis *et al*., 2009c). Because Enceladus and Titan were *a priori* formed from the same building blocks in Saturn's subnebula, the D/H ratio in Titan's water should be similar to the value measured in Enceladus. Assuming that methane was synthesized in Titan's interior, then its D/H ratio would have been very close to the D/H value of water measured in Enceladus, which is in fact about twice larger (Mousis *et al*., 2009c; Waite *et al*., 2009). This comparison rules out the mechanism of serpentinization as the main source of atmospheric methane.

The lack of detection of Xe and Kr in Titan's atmosphere motivated Mousis *et al.* (2009a) to propose a formation model of the satellite in which noble gases are trapped in clathrates formed in the protosolar nebula. In this model, temperature and pressure conditions led to the sublimation of Ar in the Saturn's subnebula, while Xe and Kr remained trapped in the methane-dominated clathrates incorporated in Titan's building blocks. This scenario relies on calculations using a statistical thermodynamic model describing the composition of clathrates formed in the protosolar nebula, which is similar to the one used to compute the planetesimals composition in Fig. 4, and adopting conditions in which argon and carbon monoxide are selectively devolatilized while methane, ammonia, krypton, and xenon are retained. Note that, to explain the Kr and Xe deficiencies in Titan's atmosphere, Mousis *et al.* (2009a) proposed that, instead of being trapped in methane-dominated clathrate, Kr and Xe could have been sequestrated by $H_3^+$ in the form of Kr-$H_3^+$ and Xe-$H_3^+$ complexes in the protosolar nebula gas phase. This hypothesis may be supported by a recent study by Pauzat *et al.* (2013) showing that the $H_3^+$ abundance is found sufficiently high in the outer part of the protosolar nebula to allow for the efficient formation of Kr-$H_3^+$ and Xe-$H_3^+$ complexes at temperatures higher than those required for their trapping in clathrates (> 50 K). Another explanation for the depletion of Kr and Xe in Titan's atmosphere was proposed by Thomas *et al.* (2007, 2008) and Mousis *et al.* (2011) who suggested that large amounts of Kr and Xe were initially available in Titan's atmosphere, and have been subsequently trapped in multiple guest clathrates dominated by nitrogen and methane and formed from the crystalline water ice available on the surface.

Methane clathrates are thought to play an important role in the present state of Titan's atmosphere. It has been proposed that, if Titan has a 100-km-thick near-surface layer of high-strength, low-thermal conductivity methane hydrate as previously suggested in the literature, then its interior is likely to be considerably warmer than previously expected (Durham *et al.*, 2003). Also, because photochemistry in the atmosphere of Titan should eliminate methane in a few tens of millions of years (Mandt *et al.*, 2009), a source of atmospheric methane is required. Observations from the Cassini-Huygens mission have ruled out large liquid bodies on the surface as a source due to their measured finite extent (Mastrogiuseppe *et al.*, 2014). Tobie *et al.* (2006) instead proposed a model where atmospheric methane is released from clathrates. Laboratory experiments have shown that methane clathrate would form easily under Titan's proto-core conditions, and would ascend to the top of the outer liquid layer. Tobie *et al.* (2006) showed that, by taking into account the coupling of Titan's orbit with its interior as well as the properties of methane clathrates, it appears that three episodes of outgassing have occurred, releasing methane into Titan's atmosphere. The last episode is supposed to have occurred after

3.5 Gyr and is enough to account for the present-day methane abundance in Titan's atmosphere.

In addition to atmospheric resupply, the dissociation of methane clathrate on Titan has also been suggested to drive explosive cryovolcanism (Lopes *et al.*, 2013).

Another potential source of clathration on Titan involves ethane. As a result of the production of this alkane in Titan's atmosphere over the Solar System's lifetime, it was expected that Cassini-Huygens would find a ~155m deep global ethane ocean on its surface, which has not been observed. Even scenarios including periodic outgassing of methane (Tobie *et al.*, 2006) and thus lower ethane production would still involve a 20m-deep global ocean. The lakes detected at high latitudes during Cassini RADAR flybys (Stofan *et al.,* 2007; Lorenz *et al.,* 2008) cannot store more than 20% of the produced ethane. Mousis and Schmitt (2008) proposed an explanation to this ethane deficiency. The different mechanisms of cryovolcanism envisioned in the case of Titan (Mitri *et al.,* 2006; Tobie *et al.,* 2006) produce a very porous material at the surface. Thus, liquid hydrocarbons are able to percolate through this layer and form a reservoir, which explains the lack of ethane at the surface. At sufficient depth, multiple guest clathrates including methane, ethane, and nitrogen can be formed. The formation of these clathrates would eventually reduce the ice porosity, isolating the hydrocarbons liquid reservoir under a depth depending on conditions. Subsequent cryovolcanic events may destabilize clathrates and break this isolation.

The presence of a layer of methane clathrates and the percolation of ethane can also be used to explain the fact that Titan is flatter than expected (polar radius 270m shorter than predicted by the flattening due to spin rate). Choukroun and Sotin (2012) proposed that percolating ethane might substitute for methane in the large cages of clathrates, as shown by laboratory experiments, and would then induce an increase of their density. Global circulation models predict that ethane produced by photolysis precipitates mainly at high latitudes (Rannou *et al*., 2006; Griffith *et al*., 2006; Mitchell *et al*., 2009), which implies that the substitution would happen at the poles and cause flattening. Choukroun and Sotin (2012) showed that the observed 270m flattening is compatible with the accepted ethane production rate over reasonable timescales. Another effect of this mechanism would be the release of methane, contributing to its presence in the atmosphere. Recently, Mousis *et al.* (2014) also considered the effects of interaction between a hydrocarbon liquid reservoir (i.e., alkanofer) and a clathrate layer (see Fig. 7). Thermodynamic equilibrium calculations show that propane and ethane compete for the same cages and that the eventual composition of the reservoir is a function of the clathrate structure that is considered. When considering a structure I clathrate forming in coexistence with the lake, the result is a liquid solution dominated by ethane if less than 0.9 mole fraction of the reservoir is trapped. Above a 0.9 mole fraction, however, propane becomes dominant. With structure II clathrate, ethane becomes the dominant constituent in the liquid phase.

## 4.2   Enceladus

Methane has been detected in the plumes emitted by Enceladus (Waite *et al.*, 2009, 2014). Mousis *et al.* (2009d) proposed a scenario for the formation of Enceladus similar to t h a t   o f the aforementioned Titan, which would trace back the origin of its methane reservoir to the same type of building blocks. Due to the position of Enceladus (nearer to Saturn), the phenomenon of devolatilization is expected to be even more pronounced than in Titan's case. However, there is no data available yet as to the abundance of Kr and Xe in Enceladus. The methane of Enceladus could alternatively be produced by hydrothermal activity such as serpentinization reactions at the interface between the rocky core and the putative internal   ocean of Enceladus (see Holm *et al.*, 2014 for details). A measurement of the D/H ratio in $CH_4$, w h i c h   i s   not possible at present, would explain its origin. Mousis *et al.* (2009d) determined that the D/H ratio would range between $4.7 \times 10^{-5}$ and $1.5 \times 10^{-4}$ in methane if it originates from the solar nebula, by taking as a reference the methane of Titan, which is primordial (Mousis *et al.*, 2009d), whereas methane produced by serpentinization reactions would have a D/H ratio between $2.1 \times 10^{-4}$ and

$4.5 \times 10^{-4}$, assuming a reaction between peridotite and deuterium-enriched residual water. The methane of Enceladus being primordial (from protosolar clathrates) or a product of hydrothermal activity would help to clarify our understanding of the subnebula's thermodynamic conditions at the time of  formation of the satellite's building blocks.

Bouquet *et al.* (2015) investigated the possibility of clathrate production in   current day Enceladus. In the plausible conditions of the putative internal ocean of Enceladus, a liquid mixture that corresponds to the composition of the plumes would produce multiple guest clathrates including methane below a 20 km depth, modifying the composition of the ocean, as shown in Fig. 8. Preliminary results show that the ultimate fate of those clathrates depends on multiple factors. For example, the low density of structure II clathrates implies that these clathrates would ascend to the top of the ocean. Depending on how thick the ice layer above the ocean is, these ices could produce a clathrate layer or dissociate and release their volatiles into the ocean. Structure I clathrates have higher density, close to that of salt water, and thus their buoyancy depends on the exact composition (and thus density) of the surrounding ocean. Thermodynamic equilibrium calculations show that methane in the ocean would be depleted by its encaging in clathrates. Therefore, studying the ultimate fate of clathrates is necessary to determine whether the abundance of methane in the plumes is due to the addition of methane by other sources (e.g., serpentinization).

## 4.3   Europa

The conditions on Europa are also favorable to clathrate formation (Kargel *et al*., 2000; Prieto-Ballesteros *et al.,* 2005; Zolotov and Kargel, 2009). Hand *et al.* (2006) studied the possibility of the radiolytically produced oxidants being trapped in clathrates. Their conclusion is that oxidants such as $CO_2$ and $SO_2$ generate a 1-10m layer of clathrates, and that the double occupancy of clathrates cages by molecular oxygen could explain the observations of solid-state oxygen at the surface (Spencer and Calvin, 2002), as these clathrates would stay stable at the surface. Such conclusions may apply to Ganymede as well.

The presence of an internal ocean on Europa raises the question of the formation of clathrates within the ocean as well the influence of the volatiles trapping on the composition of the ocean. As gravity on Europa is higher than on Enceladus, pressure conditions for clathrate formation is easily met in a cold liquid ocean. The high propensity of methane to form clathrates in the conditions of Enceladus (Bouquet *et al.*, 2015) allows us to argue that it would be trapped in clathrates in Europa's ocean as well. Depending on the exact composition of the clathrates (as they can be expected to be multi-guest), the type of clathrate structure formed, and the density of the ocean, it is possible that these clathrates would deposit at the bottom of the ocean or float and reach the bottom of the ice layer, similar to the case of Enceladus.

## 4.4   Comets and Kuiper-Belt Objects

Given their high heliocentric distances, Kuiper-Belt Objects (hereafter KBOs) have physical properties consistent with the presence of clathrates, even if ground-based observations performed so far do not allow us to determine the presence of amorphous ice versus crystalline ice on their surfaces, due to poor signal-to-noise ratios. Deep water ice absorption bands have been detected on most of large KBOs (absolute magnitude brighter than 3), while small KBOs (absolute magnitude fainter than 5) present only small absorption bands. For intermediate objects, there is only one case (2003 $AZ_{84}$) for which a deep water absorption band has been detected (Barucci *et al*., 2011; Brown *et al,*. 2012). The water ice absorption bands detected on KBOs can be fitted either with amorphous or crystalline water ice, depending on the target (Barucci *et al*., 2011). Large amounts of methane and other volatiles have been detected at the surface of some large KBOs, including Pluto and Eris (Lellouch *et al.*, 2011; Barucci *et al.*, 2011). Meanwhile, thermodynamic equilibrium calculations show that clathrate formation was possible on Pluto during its cooling phase after accretion. Formation of clathrates in equilibrium with the cooling atmosphere could have permitted the efficient trapping of the atmospheric noble gases Ar, Kr, and Xe in these crystalline structures (Mousis *et al*., 2013b). Because the propensity to trap $CH_4$ in clathrates is similar to that of Xe (Sloan and Koh, 2008), these clathrates should also incorporate some substantial fractions of $CH_4$ if they do exist on Pluto's surface or in its shallow subsurface. Given the similarities between the properties of the two bodies, the conclusions expressed for Pluto are also valid for Neptune's satellite Triton. For example, it has been proposed that clathrates could be a possible source for Triton's plumes (Croft et al. 1995).

In the case of comets, which are thought to be formed in the same outer regions of the protosolar nebula as KBOs, the presence of water ice remains difficult to detect from ground-based observations. However, the presence of crystalline water ice has been established at the surface of comet P/Tempel 1 by the Deep Impact spacecraft (Sunshine *et al*., 2006), where regions with 3 to 6% water ice particles of 10 to 50 micrometers in diameter have been directly observed. Methane has also been detected in cometary coma because of its infrared emission lines, with a production rate relative to water molecules of ~1% (Mumma and Charnley, 2011). These two observational elements are compatible with the hypothesis of a $CH_4$-rich clathrate layer that destabilizes when the comet approaches the Sun. The presence of clathrates in comets has been studied from a theoretical point of view in different papers (Marboeuf *et al.*, 2010, 2011, 2012). These authors have developed a nucleus model, which considers an initially homogeneous sphere composed of a predefined porous mixture of ices and dust and takes into account heat transmission, gas diffusion, sublimation/recondensation of volatiles within the nucleus, water ice phase transition, dust release, and mantle formation. This model has predicted that stability conditions of multiple guest clathrates are permanently present in the subsurface of the short-period comets 67P/Churyumov-Gerasimenko and 46P/Wirtanen, namely, the current and initial targets of the *Rosetta* spacecraft. The thickness of the clathrate zone slightly oscillates with time as a function of the heliocentric distance, but never vanishes.

Methane trapping by clathrates in the protosolar nebula is also thought to have influenced the present-day state of Pluto and comets (Mousis *et al*., 2012b). The thermodynamic calculations of multiple guest clathrates compositions in nebular conditions suggest that nitrogen has not been trapped in these structures, in contrast with methane and carbon monoxide. Instead, nitrogen would have condensed as pure ice at lower temperatures (~20 K) than those required for clathration (~50 K) in the protosolar nebula. This feature, associated to the fact that radiogenic heating might have been efficient in comets (Mousis *et al*., 2012b), could have induced the devolatilization of nitrogen while methane and carbon monoxide would have remained stable in the nuclei. Compared to comets, the quasi-absence of porosity and the much higher gravities in larger Kuiper-Belt Objects, Triton or Pluto would have prevented any substantial loss of nitrogen.

## 5. Conclusions

In this paper, we reviewed the possible locations and occurrence conditions of methane clathrate in the Solar System. Indeed, this clathrate is typically formed at higher temperature than many other clathrates for a given pressure and correspondingly remains stable longer against dissociation when its environment is progressively heated. It is thus expected to exist in many solid bodies of the Solar System where crystalline water ice can coexist with gaseous methane.

We first detailed the scenarios that are presently under debate for the formation conditions of $CH_4$ in the interstellar medium. Gas phase vs. surface reactions have been discussed in the literature, including formation of methane by H impacts on carbonaceous and/or icy grains of the interstellar medium. Additional information on these mechanisms is thus strongly required based, for instance, on radio observations of the methane isotopomers. Then, we have discussed the possible incorporation of methane in clathrates that could have been formed in the protosolar nebula. The corresponding mechanisms are dependent on the plausible water abundance throughout the protosolar nebula and on the fact that clathration stops when there is no more crystalline water ice available to trap the volatile species. Calculations based on accurate thermodynamic models thus showed that, beyond the snowline, methane should always be trapped in clathrates and never form pure condensates. As a consequence, methane clathrate agglomeration in comets and in the building blocks of the giant planets and their surrounding satellites is a plausible scenario for the origin of methane in solid bodies like Titan, Enceladus, Europa, comets, and Kuiper-Belt Objects. There, methane could be encaged in multi-guest clathrates together with, for instance, noble gases (Titan, Encelade), molecular oxygen (Europa and, perhaps Ganymede), CO (comets), and also $N_2$ (Triton and Pluto).

Methane clathrate is also present on Earth, originating today from biogenic sources, such as fermentation and/or anaerobic oxidation of organic material. It is formed under very specific, chemical and geological conditions and, thus, primarily occurs on the continental slopes and in the very cold Arctic regions (in the continental shelf or on the land in permafrost). Methane clathrate is attracting increasing attention because of its possible influence on global warming in case of destabilization. It also represents huge amounts of stored hydrocarbons that could be used as a nearly inexhaustible energy source in the near future.

More surprisingly, methane clathrate could also be present in Mars where a water ice cryosphere exists below the surface. Indeed, small amounts of $CH_4$ have been detected in the atmosphere of Mars, characterized by a high temporal and spatial variability. This suggests a possible release of $CH_4$, irrespective of its origin, by methane clathrates that have been shown, at least theoretically, to be stable in the martian crust, from a few to hundreds of meters below the surface. However, no direct observation of clathrates on Mars has been made to date and, concerning especially methane, both the origin of this species and its clathration/release mechanisms are still being debated.

Finally, the present review has presented scenarios for $CH_4$ enclathration in solid bodies of the Solar System. However, it should be noted that $CH_4$ clathrate is very hard to detect and most of the information detailed here is based on laboratory works (experiments and theoretical models). Significantly improved spectroscopic techniques (i.e., high spectral resolution in infrared and signal-to-noise ratio) from telescopes or space missions and the comparison of the spectra with laboratory data (Dartois, 2011; Dartois *et al.*, 2012; Oancea *et al.*, 2012) will certainly be required to directly observe methane clathrates at the surface of these solid bodies. Moreover, searching for clathrates in their subsurface will be even more challenging, and this goal is actually not reachable for a few decades. For the moment, increasing our knowledge of methane clathrates in the Solar System is still based on improvement of laboratory models (Lunine and Stevenson, 1985; Sloan and Koh, 2008; Marion et al., 2006, 2014).

**ACKNOWLEDGEMENTS. We wish** to thank Dr. Jeffrey Kargel and an anonymous Referee for their very constructive comments that helped us strengthen our manuscript. O. M. acknowledges support from CNES. E. C. acknowledges support from PNP. J. H. W. acknowledges support from Cassini-Huygens JPL subcontracts 1405851 and 1405853. The authors gratefully acknowledge support from the International Space Science Institute (ISSI) for our team "The Methane Balance - Formation and Destruction Processes on Planets, their Satellites and in the Interstellar Medium", Team ID 193. This work has been carried out thanks to the support of the A*MIDEX project (n° ANR-11-IDEX-0001-02) funded by the « Investissements d'Avenir » French Government program, managed by the French National Research Agency (ANR).

**REFERENCES**

Aikawa, Y., Herbst, E., Roberts, H., and Caselli, P. (2005) Molecular Evolution in Collapsing Prestellar Cores. III. Contraction of a Bonnor-Ebert Sphere. *Astrophys. J.* 620:330-346.

Aikawa, Y., Wakelam, V., Garrod, R. T., and Herbst, E. (2008) Molecular Evolution and Star Formation: From Prestellar Cores to Protostellar Cores. *Astrophys. J.* 674:984-996.

Aikawa, Y., and 11 colleagues (2012) AKARI observations of ice absorption bands towards edge-on young stellar objects. *Astron. Astrophys.* 538:A57.

Aikawa, Y. (2013) Interplay of Chemistry and Dynamics in the Low-Mass Star Formation. *Chem. Rev.* 113:8961-8980.

Alibert, Y., Mousis, O., and Benz, W. (2005a) On the Volatile Enrichments and Composition of Jupiter. *Astrophys J.* 622:L145-L148.

Alibert, Y., Mousis, O., Mordasini, C., and Benz, W. (2005b) New Jupiter and Saturn Formation Models Meet Observations. *Astrophys. J.* 626:L57-L60.

Ali-Dib, M., Mousis, O., Petit, J.-M., and Lunine, J.I. (2014) Carbon-rich Planet Formation in a Solar Composition Disk. *Astrophys. J.* 785:125.

Atreya, S. K., Adams, E. Y., Niemann, H. B., Demick-Montelara, J. E., Owen, T. C., Fulchignoni, M., Ferri, F., and Wilson, E. H. (2006) Titan's methane cycle. *Planet. Space Sci.* 54:1177-1187.

Atreya, S. K., Mahaffy, R. P., and Wong, A. (2007) Methane and related trace species on Mars: Origin, loss, implications for life, and habitability. *Planet. Space Sci.* 55:358-369.

Bar-Nun, A., Litman, M., and Rappaport, M. L. (1980) Interstellar molecules - Hydrocarbon formation on graphite grains at T greater than or equal to 7 K. *Astron. Astrophys.* 85:197-200.

Barucci, M.A., Alvarez-Candal, A., Merlin, F., Belskaya, I.N., de Bergh, C., Perna, D., DeMeo, F., and Fornasier, S. (2011) New insights on ices in Centaur and Transneptunian populations. *Icarus* 214:297-307.

Bibring, J.-P., Langevin, Y., Mustard, J. F., Poulet, F., Arvidson, R., Gendrin A., Gondet, B., Mangold, N., Pinet, P., and Forget, F. (2006) Global Mineralogical and Aqueous Mars History Derived from OMEGA/Mars Express Data. *Science* 312:400-404.

Bockelée-Morvan, D., Crovisier, J., Mumma, M. J., and Weaver, H. A. (2004) The composition

of cometary volatiles. *Comets II* 391-423.

Boogert, A. C. A., Schutte, W. A., Tielens, A. G. G. M., Whittet, D. C. B., Helmich, F. P., Ehrenfreund, P., Wesselius, P. R., de Graauw, T., and Prusti, T. (1996) Solid methane toward deeply embedded protostars. *Astron. Astrophys.* 315:L377-L380.

Bouquet, A., Mousis, O., Waite, J.H., and Picaud, S. (2015) Possible evidence for a methne source in Enceladus' ocean. Geophys. Res. Lett., in press, as doi: 10.1002/2014GL063013

Brown, M.E., Schaller, E.L., and Fraser, W.C. (2012) Water Ice in the Kuiper Belt. *Astron. J.* 143:146.

Buffet, B., and Archer, D. (2004) Global inventory of methane clathrate : sensitivity to changes in the deep ocean. *Earth Planet. Sci. Lett.* 227:185-199.

Carr, J. S., and Najita, J. R. (2011) Organic Molecules and Water in the Inner Disks of T Tauri Stars. *Astrophys. J.* 733:102.

Chassefière, E. (2009) Metastable methane clathrate particles as a source of methane to the Martian atmosphere. *Icarus* 204:137-144.

Chassefière, E., and Leblanc, F. (2011a) Methane release and the carbon cycle on Mars. *Planet. Space Sci.* 59:207-217.

Chassefière, E., and Leblanc, F. (2011b) Constraining methane release due to serpentinization by the D/H ratio on Mars. *Earth Planet. Sci. Lett.* 310:262–271.

Chassefière, E., Dartois, E., Herri, J.-M., Tian, F., Schmidt, F., Mousis, O., and Lakhlifi, A. (2013a) $CO_2$-$SO_2$ clathrate hydrate formation on early Mars. *Icarus* 223:878-891.

Chassefière E., Langlais B., Quesnel, Y., and Leblanc F. (2013b) The fate of early Mars' lost water: the role of serpentinization. *J. Geophys. Res.* 118:1123-1134.

Chastain, B. K., and Chevrier, V. (2007) Methane clathrate hydrates as a potential source for martian atmospheric methane. *Planetary and Space Science* 55:1246-1256.

Choukroun, M., and Sotin, C. (2012) Is Titan's shape caused by its meteorology and carbon cycle? *Geophys. Res. Let.* 39:4201.

Cruikshank, D. P., Pilcher, C. B., and Morrison, D. (1976) Pluto - Evidence for methane frost. *Science* 194:835-837.

Ciesla, F. J., and Cuzzi, J. N. (2006) The evolution of the water distribution in a viscous protoplanetary disk. *Icarus* 181:178-204.

Croft, S.K., Kargel, J. S., Kirk, R. L., Moore, J. M., Schenk, P. M., and Strom, R. G. (1995), The Geology of Triton, in *Neptune and Triton,* D.P. Cruikshank, ed., pp. 879-947, University of Arizona Press, Tucson.

Cyr, K. E., Sears, W. D., and Lunine, J. I. (1998) Distribution and Evolution of Water Ice in the Solar Nebula: Implications for Solar System Body Formation. *Icarus* 135:537-548.

Dartois, E. (2011) CO clathrate hydrate: Near to mid-IR spectroscopic signatures. *Icarus* 212: 950-956.

Dartois, E., Duret, P., Marboeuf, U., and Schmitt, B. (2012) Hydrogen sulfide clathrate hydrate FTIR spectroscopy: A help gas for clathrate formation in the Solar System? *Icarus* 220:427-434.

Dickens, G.R., O'Neil, J.R., Rea, D.K., and Owen, R.M. (1995) Dissociation of oceanic methane hydrate as a cause of the carbon isotope excursion at the end of the Paleocene. *Paleooceanography*. 10-6:965-971.

Durham, W. B., Kirby, S. H., Stern, L. A., and Zhang, W. (2003) The strength and rheology of methane clathrate hydrate. *Journal of Geophysical Research (Solid Earth)* 108:2182.

Elwood Madden, M. E., Szymcek, P., Ulrich, S. M., McCallum, S., and Phelps, T. J. (2009) Experimental formation of massive hydrate deposits from accumulation of $CH_4$ gas bubbles within synthetic and natural sediments. *Mar. Pet. Geol.* 26:369-378.

Emmanuel, S., and Ague, J. J. (2007) Implications of present-day abiogenic methane fluxes for the early Archean atmosphere. *Geophys. Res. Let.* 34, 15, CiteID L15810.

Etiope, G., Ehlmann, B.L., and Schoell, M. (2013) Low temperature production and exhalation of methane from serpentinized rocks on Earth: A potential analog for methane production on Mars. *Icarus* 224:276-285.

Fonti, S., and Marzo, G. A. (2010) Mapping the methane on Mars, *Astron. Astrophys.* 512:A51.

Formisano, V., Atreya, S., Encrenaz, T., Ignatiev, N., and Giuranna, M. (2004) Detection of methane in the atmosphere of Mars. *Science* 306:1758-1761.

Fray, N., Marboeuf, U., Brissaud, O., and Schmitt, B. (2010) Equilibrium Data of Methane, Carbon Dioxide, and Xenon Clathrate Hydrates below the Freezing Point of Water. Applications to Astrophysical Environments, *J. Chem. Eng. Data*, 55(11):5101-5108.

Gautier, D., Hersant, F., Mousis, O., and Lunine, J.I. (2001) Enrichments in Volatiles in Jupiter: A New Interpretation of the Galileo Measurements. *Astrophys. J.* 550:L227-L230.

Geminale, A., Formisano, V., and Sindoni, G. (2011) Mapping methane in Martian atmosphere

with PFS-MEX data. *Planet. Space Sci.* 59:137-148.

Gough, R.V., Turley, J.J., Ferrell, G.R., Cordova, K.E., Wood, S.E., Dehaan, D. O., McKay, C. P., Toon, O. B., and Tolbert, M. A. (2011) Can rapid loss and high variability of Martian methane be explained by surface $H_2O_2$? *Planet. Space Sci.* 59:238-246.

Griffith, C.A., and 13 colleagues (2006) Evidence for a Polar Ethane Cloud on Titan. *Science* 313:1620-1622.

Grossman, L. (1972) Condensation in the primitive solar nebula. *Geochimica et Cosmochimica Acta* 36:597-619.

Hand, K. P., Carlson, R. W., Cooper, J., and Chyba, C. F. (2006) Clathrate hydrates of oxidants in the ice shell of Europa. *Astrobiology* 6(3):463-482.

Herbst, E., and Klemperer, W. (1973) The Formation and Depletion of Molecules in Dense Interstellar Clouds. *Astrophys. J.* 185:505-534.

Herri, J.-M., and Chassefière, E. (2012) Carbon dioxide, argon, nitrogen and methane clathrate hydrates: thermodynamic modelling, investigation of their stability in Martian atmospheric conditions and variability of methane trapping. *Planet. Space Sci.* 73:376-386.

Higgins, J.A., and Schrag, D.P. (2006) Beyond methane: Towards a theory for the Paleocene–Eocene Thermal Maximum. *Earth Planet. Sci. Lett.* 245:523-537.

Hogerheijde, M. R., and 14 colleagues (2011) Detection of the Water Reservoir in a Forming Planetary System. *Science* 334:338.

Holm, N. et al. (2015) Serpentinization on planets, moons and comets. *Astrobiology*, submitted.

Howarth, R. W., Santoro, R., and Ingraffea, A. (2011) Methane and the greenhouse-gas footprint of natural gas from shale formations, Climatic Change, DOI 10.1007/s10584- 011-0061-5.

Irvine, W.M. (1987) The chemistry of cold, dark interstellar clouds. *Astrochemistry* 120:245-251.

Irwin, P. G. J. (2003). Giant planets of our solar system: atmospheres compositions, and structure. Springer Praxis books in geophysical sciences. Berlin: Springer.

Kaminska, M., Zhaunerchyk, V., Vigren, E., Danielsson, M., Hamberg, M., Geppert, W. D., Larsson, M., Rosen, S., Thomas, R. D., and Semaniak, J. (2010) Dissociative recombination of $CH_5^+$ and $CD_5^+$: Measurement of the product branching fractions and the absolute cross sections, and the breakup dynamics in the $CH_3 + H + H$ product channel. *Phys. Rev. A* 81:062701.

Kargel, J. S., and Lunine, J. I. (1998) Clathrate hydrates on Earth and in the solar system, *Solar System Ices*, de Bergh, C., Festou, M., and Schmitt, B. (Eds.), pp 97-117, Kluwer Academic Publishers.

Kargel, J. S., Tanaka, K. L., Baker, V. R., Komatsu, G., Macayeal, D. R. (2000) Formation and Dissociation of Clathrate Hydrates on Mars: Polar Caps, Northern Plains and Highlands. *Lunar and Planetary Science Conference* 31, 1891.

Kargel, J. S., Kaye, J. Z., Head, J. W., Marion, G. M., Sassen, R., Crowley, J. K., Ballesteros, O. P., Grant, S. A., and Hogenboom, D. L. (2000) Europa's Crust and Ocean: Origin, Composition, and the Prospects for Life. *Icarus* 148:226-265.

Kargel, J.S. (2004) *Mars: A Warmer Wetter Planet*, Praxis-Springer (publ.), 557 pp., ISBN: 978-1-85233-568-7.

Kargel, J. S., Furfaro, R., Prieto-Ballesteros, O., Rodriguez, J. A. P., Montgomery, D. R., Gillespie, A. R., Marion, G. M., and Wood, S. E. (2007) Martian hydrogeology sustained by thermally insulating gas and salt hydrates. *Geology* 35(11):975-978.

Kasting, J.F. (1997) Warming early Earth and Mars. *Science* 226:1213-1215.

Kennett, J. P., Cannariato, K. G., Hendy, I. L., and Behl, R. J. (2000) Carbon Isotopic Evidence for Methane Hydrate Instability During Quaternary Interstadials. *Science* 288 (5463):128-133.

Kennett, J. P., Cannariato, K. G., Hendy, I. L., Behl, R. J. (2003) Methane Hydrates in Quaternary Climate Change: The Clathrate Gun Hypothesis. Washington DC: American Geophysical Union. ISBN 0-87590-296-0.

Knez, C., Lacy, J. H., Evans, N. J., van Dishoeck, E. F., and Richter, M. J. (2009) High-Resolution Mid-Infrared Spectroscopy of NGC 7538 IRS 1: Probing Chemistry in a Massive Young Stellar Object. *Astrophys. J.* 696:471-483.

Kouchi, A., Yamamoto, T., Kozasa, T., Kuroda, T., and Greenberg, J. M. (1994) Conditions for condensation and preservation of amorphous ice and crystallinity of astrophysical ices. *Astron. Astrophys.* 290:1009-1018.

Krasnopolsky, V. A., Maillard, J.-P. and Owen, T. C. (2004) Detection of methane in the Martian atmosphere: evidence for life? *Icarus* 172:537-547.

Krasnopolsky, V.A. (2012) Search for methane and upper limits to ethane and $SO_2$ on Mars. *Icarus* 217:144-152.

Kvenvolden, K.A. (1998) A primer on the geological occurrence of gas hydrate. In Gas Hydrates: Relevance to World Margin Stability and Climatic Change (eds. J.-P. Henriet, J. Mienert),

Geological Society Special Publication No. 137, The Geological Society Publishing House, London, UK, p. 9-30.

Lacy, J. H., Carr, J. S., Evans, N. J., II, Baas, F., Achtermann, J. M., and Arens, J. F. (1991) Discovery of interstellar methane - Observations of gaseous and solid $CH_4$ absorption toward young stars in molecular clouds. *Astrophys. J.* 376:556-560.

Langlais, B., Purucker, M.E., and Mandea, M. (2004) Crustal magnetic field of Mars. *J. Geophys. Res.* 109, doi:10.1029/2003JE002048.

Lasue, J., Mangold, N., Hauber, E., Clifford, S., Feldmann, W., Gasnault, O., Grima, C., Maurice, S., and Mousis, O. (2013) Quantitative assessments of the Martian hydrosphere. *Space Sci. Rev.* 174:155-212.

Lasue, J., Langlais, B., Quesnel, Y., and Chassefière, E. (2014) Methane Trapping Capacity of the Early Martian Cryosphere, *The Eight International Conference on Mars*, July 14-18, 2014, Pasadena, California.

Lefèvre, F., and Forget, F. (2009) Observed variations of methane on Mars unexplained by known atmospheric chemistry and physics. *Nature* 460:720-723.

Lellouch, E., de Bergh, C., Sicardy, B., Käufl, H.U., and Smette, A. (2011) High resolution spectroscopy of Pluto's atmosphere: detection of the 2.3 mm $CH_4$ bands and evidence for carbon monoxide. *Astron. Astrophys.* 530:L4.

Lewis, J. S. (1972) Low Temperature Condensation from the Solar Nebula. *Icarus* 16:241- 252.

Lewis, J. S., Prinn, R. G. (1980) Kinetic inhibition of CO and N2 reduction in the solar nebula. *The Astrophysical Journal* 238:357-364.

Lopes, R. M. C., and 15 colleagues (2013) Cryovolcanism on Titan: New results from Cassini RADAR and VIMS. *Journal of Geophysical Research (Planets)* 118:416-435.

Lorenz, R.D., and 15 colleagues (2008) Fluvial channels on Titan: Initial Cassini RADAR observations. *Planet. Space Sci.* 56:1132-1144.

Lunine, J.I., and Stevenson, D.J. (1985) Thermodynamics of clathrate hydrate at low and high pressures with application to the outer solar system. *Astrophys. J. Sup. Series* 58:493- 531.

Lunine, J. I. (2006) Origin of Water Ice in the Solar System. Meteorites and the Early Solar System II 309-319.

Lyons, J., Manning, C., and Nimmo, F. (2005) Formation of methane on Mars by fluid-rock interaction in the crust. *Geophys. Res Let.* 32, 13, L13201.1-L13201.4.

Mahaffy, P.R., Webster, C.R., Atreya, S.K., Franz, H., Wong, M., Conrad, P.G. , Harpold, D., Jones, J.J., Leshin, L.A., Manning, H., Owen, T., Pepin, R.O., Squyres, S., Trainer, M. and MSL Science Team (2013) Abundance and isotopic composition of gases in the Martian atmosphere from the Curiosity rover. *Science* 341:263-265.

Mandt, K.E., Waite, J.H. Jr., Lewis, W., Magee, B., Bell, J., Lunine, J.I., Mousis, O., and Cordier, D. (2009) Isotopic evolution of the major constituents of Titan's atmosphere based on Cassini data. *Planet. Space Science* 57:1917-1930.

Marboeuf, U., Mousis, O., Petit, J.-M., and Schmitt, B. (2010) Clathrate Hydrates Formation in Short-Period Comets. *Astrophys. J.* 708:812-816.

Marboeuf, U., Mousis, O., Petit, J.-M., Schmitt, B., Cochran, A.L., and Weaver, H.A. (2011) On the stability of clathrate hydrates in comets 67P/Churyumov-Gerasimenko and 46P/Wirtanen. *Astron. Astrophys.* 525:A144.

Marboeuf, U., Schmitt, B., Petit, J.-M., Mousis, O., and Fray, N. (2012) A cometary nucleus model taking into account all phase changes of water ice: amorphous, crystalline, and clathrate. *Astron. Astrophys.* 542:A82.

Marion, G.M., Catling, D. C., and Kargel, J. S. (2006) Modeling gas hydrate equilibria in electrolyte solutions, *Computer Modeling of Phase Diagrams and Thermochemistry* 30, 248-259.

Marion, G. M., Kargel, J. S., Catling, D. C., and Lunine, J. I. (2014) Modeling nitrogen-gas, -liquid, -solid chemistries at low temperatures (173–298 K) with applications to Titan, *Icarus* 236:1-8.

Mastrogiuseppe, M., Poggiali, V., Hayes, A., Lorenz, R., Lunine, J.I., Picardi, G., Seu, R., Flamini, E., Mitri, G., Notarnicolas, C., Paillou, P., and Zebker, H (2014) The bathymetry of a Titan sea. *Geophys. Res. Let.* 41:1432-1437.

Max, M. D., Clifford, S. M. (2000) The state, potential distribution, and biological implications of methane in the Martian crust. *Journal of Geophysical Research* 105, 4165- 4172.

Meslin, P.-Y., Gough, R., Lefèvre, F. and Forget, F. (2011) Little variability of methane on Mars induced by adsorption in the regolith. *Planet. Space Sci.* 59:247-258.

Millar, T. J., Farquhar, P. R. A., and Willacy, K. (1997) The UMIST Database for Astrochemistry 1995. *Astron. Astrophys. Sup. Ser.* 121:139-185.

Mitchell, J.L., Pierrehumbert, R.T., Frierson, D.M.W., and Caballero, R. (2009) The impact of methane thermodynamics on seasonal convection and circulation in a model Titan atmosphere. *Icarus* 203:250-264.

Mitri, G., Showman, A.P., Lunine, J.I., and Lopes, R. (2006) Resurfacing of Titan by Ammonia-Water Cryomagma. *37th An. Lun. Planet. Sci. Conf.* 37:1994.

Mousis, O., Gautier, D., Bockelée-Morvan, D., Robert, F., Dubrulle, B., and Drouart, A. (2000) Constraints on the Formation of Comets from D/H Ratios Measured in $H_2O$ and HCN. *Icarus* 148:513-525.

Mousis, O., Gautier, D., Bockelée-Morvan, D. (2002a) An evolutionary turbulent model of Saturn's subnebula: implications for the origin of the atmosphere of Titan. *Icarus* 156:162- 175.

Mousis, O., Gautier, D., and Coustenis, A. (2002b) The D/H Ratio in Methane in Titan: Origin and History. *Icarus* 159:156-165.

Mousis, O. (2004) Modeling the thermodynamical conditions in the Uranian subnebula - Implications for regular satellite composition. *Astron. Astrophys.* 413:373-380.

Mousis, O., and Gautier, D. (2004) Constraints on the presence of volatiles in Ganymede and Callisto from an evolutionary turbulent model of the Jovian subnebula. *Planet. Space Sci.* 52:361-370.

Mousis, O., and Schmitt, B. 2008. Sequestration of Ethane in the Cryovolcanic Subsurface of Titan. *Astrophys. J.* 677:L67-L70.

Mousis, O., Lunine, J.I., Thomas, C., Pasek , M., Marboeuf, U., Alibert, Y., Ballenegger, V., Cordier, D., Ellinger, Y., Pauzat, F., and Picaud, S. (2009a) Clathration of Volatiles in the Solar Nebula and Implications for the Origin of Titan's Atmosphere. *Astrophys. J.* 691:1780-1786.

Mousis, O., Marboeuf, U., Lunine, J.I., Alibert, Y., Fletcher, L.N., Orton, G.S., Pauzat, F., and Ellinger, Y. (2009b) Determination of the Minimum Masses of Heavy Elements in the Envelopes of Jupiter and Saturn. *Astrophys. J.* 696:1348-1354.

Mousis, O., Lunine, J.I., Pasek, M., Cordier, D., Hunter Waite, J., Mandt, K.E., Lewis, W.S., and Nguyen, M.-J. (2009c) A primordial origin for the atmospheric methane of Saturn's moon Titan. *Icarus* 204:749-751.

Mousis, O., Lunine, J.I., Waite, J.H., Magee, B., Lewis, W.S., Mandt, K.E., Marquer, D., and Cordier, D. (2009d) Formation Conditions of Enceladus and Origin of Its Methane Reservoir. *Astrophys J.* 701:L39-L42.

Mousis, O., Lunine, J.I., Picaud, S., and Cordier, D. (2010) Volatile inventories in clathrate hydrates formed in the primordial nebula. *Farad. Disc.* 147:509.

Mousis, O., Lunine, J.I., Picaud, S., Cordier, D., Waite, J.H., Jr., and Mandt, K.E. (2011) Removal of Titan's Atmospheric Noble Gases by Their Sequestration in Surface Clathrates. *Astrophys. J.* 740:L9.

Mousis, O., Lunine, J.I., Madhusudhan, N., and Johnson, T.V. (2012a) Nebular Water Depletion as the Cause of Jupiter's Low Oxygen Abundance. *Astrophys J.* 751, L7.

Mousis, O., Guilbert-Lepoutre, A., Lunine, J.I., Cochran, A.L., Waite, J.H., Petit, J.-M., and Rousselot, P. (2012b). The Dual Origin of the Nitrogen Deficiency in Comets: Selective Volatile Trapping in the Nebula and Postaccretion Radiogenic Heating. *Astrophys. J.* 757:146.

Mousis, O., Lunine, J.I., Chassefière, E., Montmessin, F., Lakhlifi, A., Picaud, S., Petit, J.-M. and Cordier, D (2012c) Mars cryosphere : a potential reservoir for heavy noble gases? *Icarus* 218:80-87.

Mousis, O., Chassefière, E., Lasue, J., Chevrier, V., Elwood Madden, L.E., Lakhlifi, A., Lunine, J.I., Montmessin, F., Picaud, Schmidt, F., and Swindle, T.D. (2013a) Volatile trapping in Martian clathrates. *Space Sci. Rev.* 174(1-4):213-250.

Mousis, O., Lunine, J.I., Mandt, K.E., Schindhelm, E., Weaver, H.A., Alan Stern, S., Hunter

Waite, J., Gladstone, R., and Moudens, A. (2013b) On the possible noble gas deficiency of Pluto's atmosphere. *Icarus* 225:856-861.

Mousis, O., Lakhlifi, A., Picaud, S., Pasek, M., and Chassefière, E. (2013c) On the Abundances of Noble and Biologically Relevant Gases in Lake Vostok, Antarctica. *Astrobiology* 13:380-390.

Mousis, O., Choukroun, M., Lunine, J.I., and Sotin, C. (2014) Equilibrium composition between liquid and clathrate reservoirs on Titan. *Icarus* 239:39-45.

Mumma, M. J., Villanueva, G. L., Novak, R. E., Hewagama, T., Bonev, B. P., DiSanti, M. A., Mandell, A. M. and Smith, D. M. (2009) Strong Release of Methane on Mars in Northern Summer 2003. *Science* 323:1041-1045.

Mumma, M.J., and Charnley, S.B. (2011) The Chemical Composition of Comet - Emerging Taxonomies and Natal Heritage. *An. Rev. Astron. Astrophys.* 49:471-524.

Notesco, G., and Bar-Nun, A. (2005) A ~25 K temperature of formation for the submicron ice grains which formed comets. *Icarus* 175:546-550.

Oancea, A., Grasset, O., Le Menn, E., Bollengier, O., Bezacier, L., Le Mouélic, S., and Tobie, G. (2012) Laboratory infrared reflection spectrum of carbon dioxide clathrate hydrates for astrophysical remote sensing applications. *Icarus* 221:900-910.

Oberg, K. I., Boogert, A. C. A., Pontoppidan, K. M., Blake, G. A., Evans, N. J., Lahuis, F., and van Dishoeck, E. F. (2008) The c2d Spitzer Spectroscopic Survey of Ices around Low-Mass Young Stellar Objects. III. CH4. *Astrophys. J.* 678:1032-1041.

Oppenheimer, B. R., Kulkarni, S. R., Matthews, K., and Nakajima, T. (1995) Infrared Spectrum of the Cool Brown Dwarf Gl 229B. *Science* 270:1478-1479.

Owen, T., and Cess, R. D. (1975) Methane absorption in the visible spectra of the outer planets and Titan. *Astrophys. J.* 197:L37-L40.

Owen, T., Biemann, K., Rushneck, D.R., Biller, J.E., Howarth, D.W. and LaFleur, A.L. (1977) The composition of the atmosphere at the surface of Mars. *J. Geophys. Res.* 82:4635-4639.

Owen, T., Maillard, J.P., DeBergh, C., and Lutz, B.L. (1988) Deuterium on Mars: the abundance of HDO and the value of D/H. *Science* 240:1767-1770.

Owen, T., Mahaffy, P., Niemann, H.B., Atreya, S., Donahue, T., Bar-Nun, A., and de Pater, I. (1999) A low-temperature origin for the planetesimals that formed Jupiter. *Nature* 402:269-270.

Oze, C., and Sharma, M. (2005) Have olivine, will gas : Serpentinization and the abiogenic production of methane on Mars. *Geophys. Res. Lett.* 32, L10203.

Pauzat, F., Ellinger, Y., Mousis, O., Ali-Dib, M., and Ozgurel, O. (2013) Gas-phase Sequestration of Noble Gases in the Protosolar Nebula: Possible Consequences on the Outer Solar System Composition. *Astrophys. J.* 777:29.

Pellenbarg, R. E., Max, M. D., Clifford, S. M. (2003) Methane and carbon dioxide hydrates on Mars: Potential origins, distribution, detection, and implications for future in situ resource utilization. *Journal of Geophysical Research (Planets)* 108:8042.

Pontoppidan, K. M., Dullemond, C. P., van Dishoeck, E. F., Blake, G. A., Boogert, A. C. A., Evans, N. J., II, Kessler-Silacci, J. E., and Lahuis, F. (2005) Ices in the Edge-on Disk CRBR 2422.8-3423: Spitzer Spectroscopy and Monte Carlo Radiative Transfer Modeling. *Astrophys. J.* 622:463-481.

Prasad, S. S., and Huntress, W. T., Jr. (1980) A model for gas phase chemistry in interstellar clouds. I - The basic model, library of chemical reactions, and chemistry among C, N, and O compounds. *Astrophys. J. Sup. Ser.* 43:1-35.

Prieto-Ballesteros, O., Kargel, J. S., Fernandez-Sampedro, M., Selsis, F., Martinez, E. S., and Hogenboom, D. L. (2005) Evaluation of the possible presence of clathrate hydrates in Europa's icy shell or seafloor. *Icarus* 177:491-505.

Prieto-Ballesteros, O., Kargel, J. S., Fairén, A. G., Fernandez-Remolar, D. C., Dohm, J. M., and Amils, R. (2006) Interglacial clathrate destabilization on Mars: Possible contributing source of its atmospheric methane. *Geology* 34:149-152.

Prinn, R. G. P., Fegley, B., Jr. (1989) Solar nebula chemistry: origins of planetary, satellite and cometary volatiles. Origin and Evolution of Planetary and Satellite Atmospheres 78- 136.

Rannou, P., Montmessin, F., Hourdin, F., and Lebonnois, S. (2006) The Latitudinal Distribution of Clouds on Titan. *Science* 311:201-205.

Renssen, H., Beets, C. J., Fichefet, T., Goosse, H., and Kroon, D. (2004) Modeling the climate response to a massive methane release from gas hydrates. *Paleoceanography* 19, PA2010, doi:10.1029/2003PA000968.

Ross, R.G., and Kargel, J.S. (1998) Thermal conductivity of solar system ices, with special reference to Martian polar caps, *Solar System Ices*, de Bergh, C., Festou, M., and Schmitt, B. (Eds.), pp 33-62, Kluwer Academic.

Sakai, N., Sakai, T., Hirota, T., and Yamamoto, S. (2008) Abundant Carbon-Chain Molecules toward the Low-Mass Protostar IRAS 04368+2557 in L1527. *Astrophys. J.* 672:371-381.

Sakai, N., Sakai, T., Hirota, T., and Yamamoto, S. (2009) Deuterated Molecules in Warm Carbon Chain Chemistry: The L1527 Case *Astrophysical J.* 702:1025

Sauter, E.J., Muyakshin, S.I., Charlou J.-L., Schlüter, M., Boetius, A., Jerosch, K., Damm, E., Foucher, J.-P., and Klages, M. (2006) Methane discharge from a deep-sea submarine mud volcano into the upper water column by gas hydrate-coated methane bubbles. *Earth Planet. Sci. Lett.* 243:354-365.

Semaniak, J., Larson, A., Le Padellec, A., Stroemholm, C., Larsson, M., Rosen, S., Peverall, R., Danared, H., Djuric, N., Dunn, G.H., and Datz, S. (1998) Dissociative recombination and excitation of $CH_5^+$: Absolute cross sections and branching fractions *Astrophys. J.* 498: 886–895.

Schaefer, H., Whiticar, M. J., Brook, E. J., Petrenko, V. V., and Ferretti, D. F., and Severinghaus, J. P. (2006) Ice record of delta13C for atmospheric CH4 across the Younger Dryas-Preboreal transition. Science 313 (5790): 1109-12.

Sexton, P.F., Norris, R.D., Wilson, P.A., Pälike, H., Westerhold, T., Röhl, U., Bolton, C.T., and Gibbs, S. (2011) Eocene global warming events driven by ventilation of oceanic dissolved organic carbon. *Nature*. 471:349-353.

Sloan, E.D., and Koh, C.A. (2008) Clathrate hydrates of natural gases, third ed. CRC Press, Taylor & Francis Group, Boca Raton.

Sowers, T. (2006) Late quaternary atmospheric CH4 isotope record suggests marine clathrates are stable. Science 311 (5762): 838-840.

Stansberry, J. A., Spencer, J. R., Schmitt, B., Benchkoura, A. I., Yelle, R. V., and Lunine, J. I. (1996) A model for the overabundance of methane in the atmospheres of Pluto and Triton. *Planet. Space Sci*. 44:1051-1063.

Sunshine, J.M., and 22 colleagues (2006) Exposed Water Ice Deposits on the Surface of Comet 9P/Tempel 1. *Science* 311:1453-1455.

Spencer, J.R., and Calvin, W.M. (2002) Condensed $O_2$ on Europa and Callisto. *Astron. J.* 124:3400-3403.

Stern, L.A., Circone, S., Kirby, S.H., and Durham, W.B. (2001) Anomalous preservation of pure methane hydrate at 1 atm.. *J. Phys. Chem. B* 105:1756-1762.

Stevenson, D.J., and Lunine, J.I. (1988) Rapid formation of Jupiter by diffuse redistribution of water vapor in the solar nebula. *Icarus* 75:146-155.

Stofan, E.R., and 37 colleagues (2007) The lakes of Titan. *Nature* 445:61-64.

Svensen, H., Planke, S., Malthe-Sørenssen, A., Jamtveit, B., Myklebust, R., Eidem, T.R., and Rey, S.S. (2004) Release of methane from a volcanic basin as a mechanism for initial Eocene global warming. *Nature*. 429:542-545

Talbi, D., and Bacchus-Montabonel, M. C. (1998) Ab-initio study of a radiative association mechanism application to the $CH_3^+ + H_2$ reaction. *Chem. Phys*. 232:267-273.

Terada, H., Tokunaga, A. T., Kobayashi, N., Takato, N., Hayano, Y., and Takami, H. (2007) Detection of Water Ice in Edge-on Protoplanetary Disks: HK Tauri B and HV Tauri   C. *Astrophys. J*. 667:303-307.

Thi, W. F., Pontoppidan, K. M., van Dishoeck, E. F., Dartois, E., and d'Hendecourt, L. (2002) Detection of abundant solid CO in the disk around CRBR 2422.8-3423. *Astron. Astrophys*. 394:L27-L30.

Thomas, C., Mousis, O., Ballenegger, V., and Picaud, S. (2007) Clathrate hydrates as a sink of noble gases in Titan's atmosphere. *Astron. Astrophys*. 474:L17-L20.

Thomas, C., Picaud, S., Mousis, O., and Ballenegger, V. (2008) A theoretical investigation into the trapping of noble gases by clathrates on Titan. *Planet. Space Sci*. 56:1607-1617.

Thomas, C., Mousis, O., Picaud, S. and Ballenegger, V. (2009) Variability of the methane traping in martian subsurface clathrate hydrates. *Planet. Space Sci*. 57:42-47.

Tobie, G., Lunine, J.I., and Sotin, C. (2006) Episodic outgassing as the origin of atmospheric methane on Titan. *Nature* 440:61-64.

Villanueva, G.L., Mumma, M.J., Novak, R.E., Radeva, Y.L., Käufl, H.U., Smette, A., Tokunaga, A., Khayat, A., Encrenaz, T., and Hartogh, P. (2013) A sensitive search for organics ($CH_4$, $CH_3OH$, $H_2CO$, $C_2H_6$, $C_2H_2$, $C_2H_4$), hydroperoxyl ($HO_2$), nitrogen compounds ($N_2O$, $NH_3$, $HCN$) and chlorine species ($HCl$, $CH_3Cl$) on Mars using ground-based high-resolution infrared spectroscopy. *Icarus* 223:11-27.

Visscher, C., Moses, J. I. (2011) Quenching of Carbon Monoxide and Methane in the Atmospheres of Cool Brown Dwarfs and Hot Jupiters. *Astrophys. J.* 738, 72.

Waite, J. H., Jr., and 15 colleagues (2009) Liquid water on Enceladus from observations of ammonia and $^{40}$Ar in the plume. *Nature* 460:487-490.

Waite, J. H., Brockwell, T., Lewis, W. S., Magee, B., McKinnon, W. B., Mousis, O., Bouquet, A. (2014) Enceladus Plume Composition. *LPI Contributions* 1774:4013.

Webster, C.R., Mahaffy, P.R., Flesh, G.J., Niles, P.B., Jones, J.H., Leshin, L.A., Atreya, S.K., Stern, J.C., Christensen, L.E., Owen, T., Frantz, H., Pepin, R.O., Steel, A., and MSL Science Team (2013) Isotope ratios of H, C and O in $CO_2$ and $H_2O$ of the Martian atmosphere. *Science* 341:260-263.

Webster, C.R., Mahaffy, P.R., Atreya, S.K., Flesh, G.J., Mischna, M.A., Meslin, P.-Y., Farley, K.A., Conrad, P.G., Christensen, L.E., Pavlov, A.A., Martin-Torres, J., Zorzano, M.-P., McConnochie, T.H., Owen, T., Eigenbrode, J.L., Glavin, D.P., Steele, A., Malespin, C.A., Archer Jr, P.D., Sutter, B., Coll, P., Freissinet, C., McKay, C.P., Moores, J.E., Schwenzer, S.P., Bridges, J.C., Navarro-Gonzalez, R., Gellert, R., Lemmon, M.T., and the MSL Science Team (2015) Mars methane detection and variability at Gale Crater. *Science* 347:415-417.

Woodall, J., Agundez, M., Markwick-Kemper, A. J., and Millar, T. J. (2007) The UMIST database for astrochemistry 2006. *Astron Astrophys* 466:1197-1204.

Zahnle, K., Freedman, R.S., and Catling, D.C. (2011) Is there methane on Mars? *Icarus* 212:493-503.

Zhang, K., Pontoppidan, K. M., Salyk, C., and Blake, G. A. (2013) Evidence for a Snow Line beyond the Transitional Radius in the TW Hya Protoplanetary Disk. *Astrophys. J.* 766:82.

Zolotov, M. Yu., and Kargel, J. S. (2009) On the Composition of Europa's Icy Shell, Ocean and Underlying Rocks. In *Europa* (R. Pappalardo, W.B. McKinnon, and K. Khurana, eds.). Univ. of Arizona Press, Tucson, 431-457.

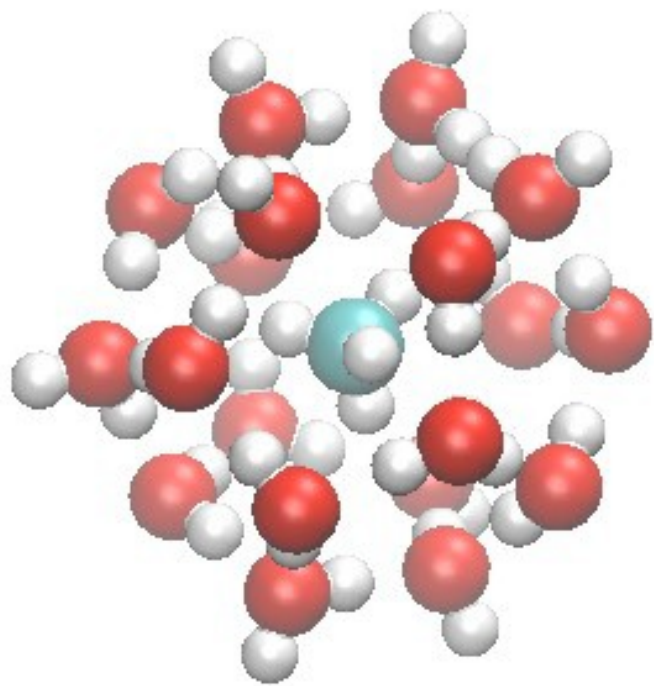

**Figure 1.** Methane clathrate. Small cage is shown.

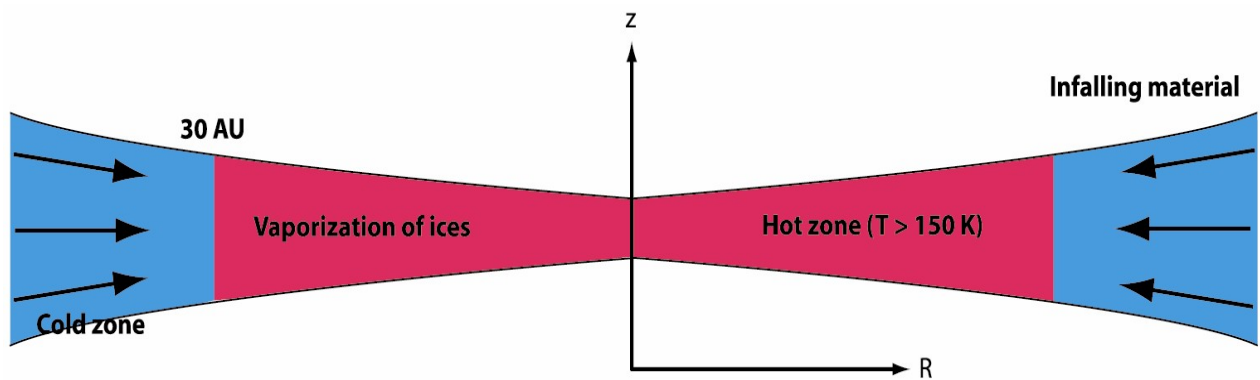

**Figure 2.** Two reservoirs of ices in the protosolar nebula. A first reservoir (cold zone), located at distances higher than ~30 AU, is constituted from amorphous ices coming from interstellar medium. The second reservoir (hot zone), located within the ~30 AU, is made from volatiles initially in the form of amorphous ices that were transported from interstellar medium towards the inner and hot part of the disk. When reaching regions with temperatures higher than ~150 K, these ices vaporized. During the cooling of the disk, volatiles located in the inner 30 AU condensed again but in crystalline forms, including both pure crystalline ices and clathrates (see text).

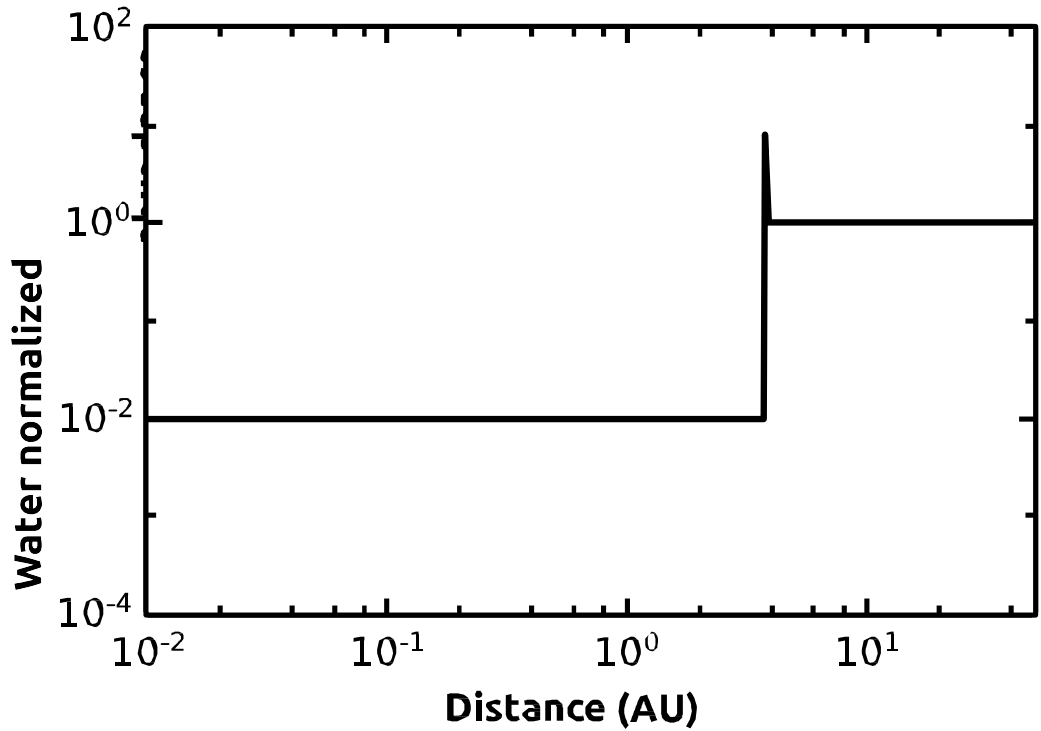

**Figure 3.** Water abundance profile throughout the protosolar nebula (normalized relative to the protosolar oxygen abundance, and assuming that all O forms $H_2O$ in the disk). Water is highly depleted at distances closer than the snowline location in the protosolar nebula. The water abundance becomes supersolar at the snowline location and remains protosolar at higher heliocentric distance. We refer the reader to Ali-Dib et al. (2014) for details about the model used in this computation.

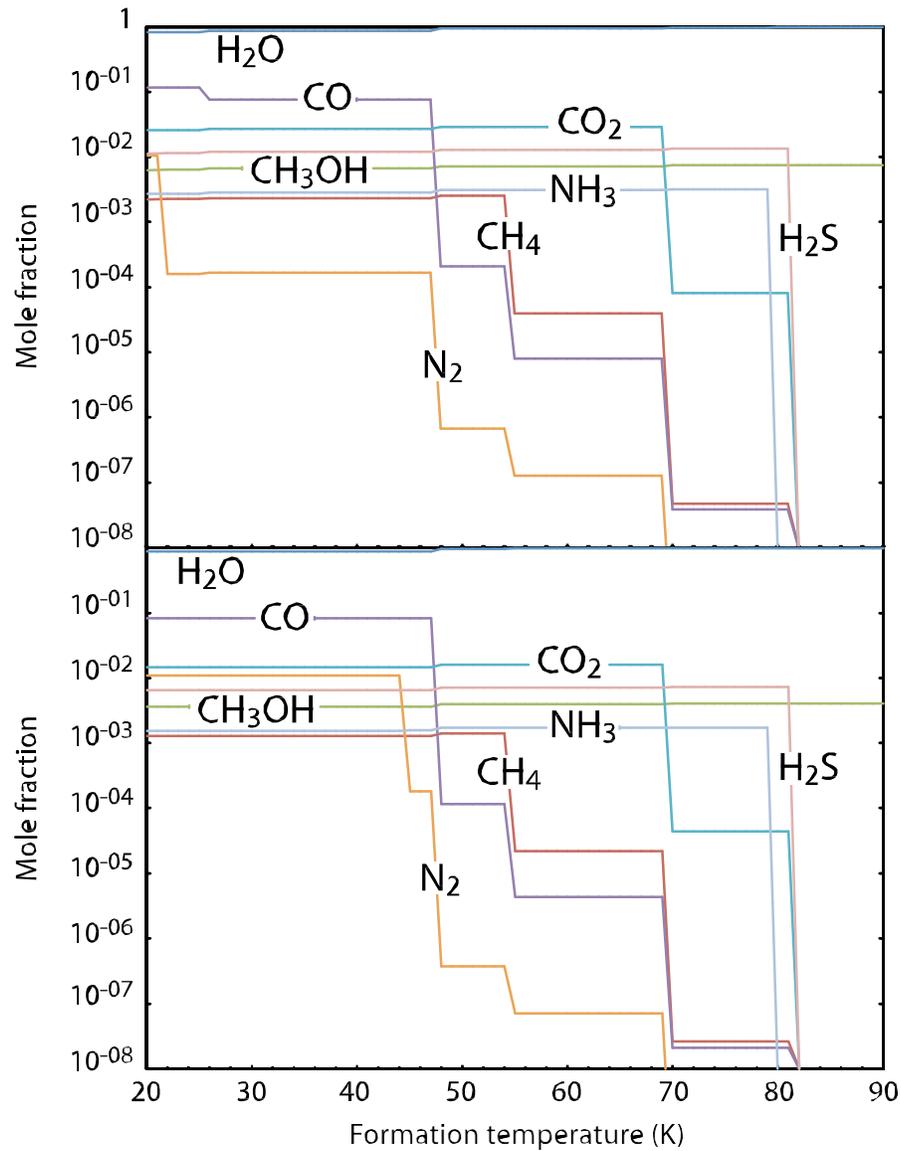

**Figure 4:** Composition of icy planetesimals as a function of their formation temperature in the protosolar nebula (see Mousis *et al.*, 2012b for details about the model). Top panel: the water abundance in the protosolar nebula is derived from protosolar values for oxygen and carbon, assuming that all O, C and N are in the forms of $H_2O$, CO, $CO_2$, $CH_3OH$ and $CH_4$, $N_2$ and $NH_3$, with CO:$CO_2$:$CH_3OH$:$CH_4$ = 70:10:2:1 and $N_2$:$NH_3$ = 10:1 in the disk gas phase (Johnson *et al.*, 2012). In this case, water is not abundant enough to form $NH_3$-$H_2O$ hydrate and trap all the other volatiles in multiple guest clathrates. If the protosolar nebula cools down to temperatures lower

than ~25 K, then $N_2$ essentially crystallizes as pure ice, as well as a significant fraction of CO. Bottom panel: the water abundance is arbitrarily set 30% higher than the protosolar oxygen abundance, namely, the minimum abundance from which $NH_3$-$H_2O$ hydrate forms and all the other volatiles are trapped in multiple guest clathrates during the cooling of the protosolar nebula. The temperature of volatiles encaging always remains higher than ~45 K at nebular pressure conditions.

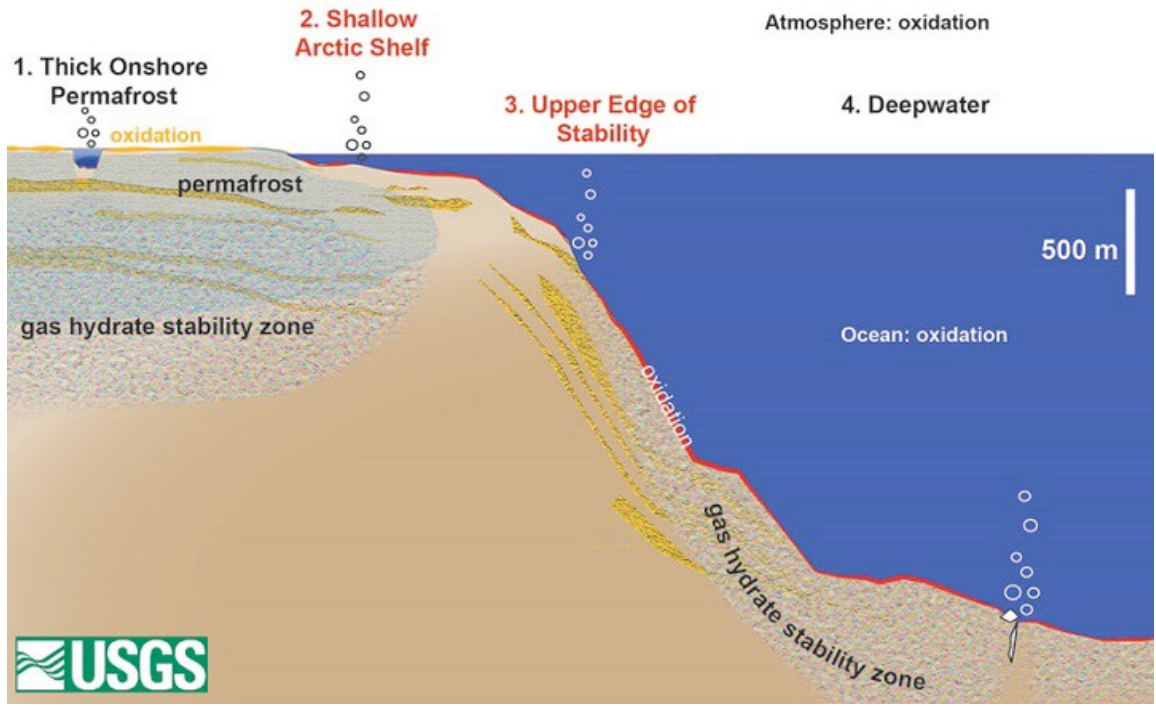

**Figure 5:** Methane clathrate deposits near Artic coast (credit USGS). Because of the future rise of sea levels and global temperature, clathrates present in the permafrost and at shallow depth in the oceans might destabilize and release the trapped methane into the atmosphere.

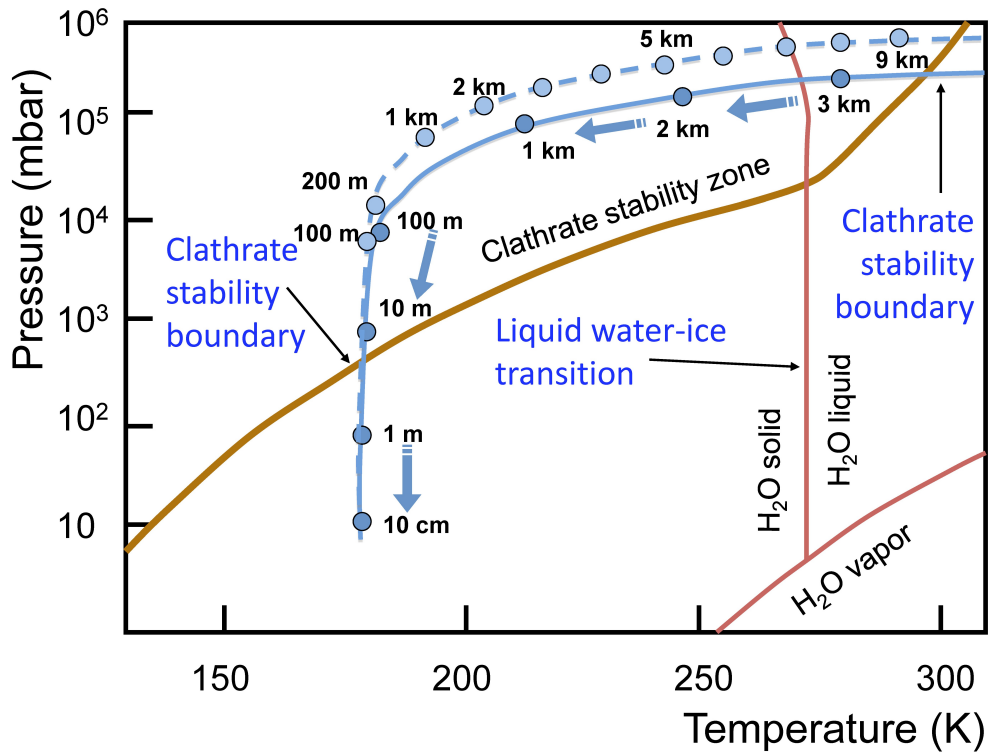

**Figure 6:** Phase diagram of CH₄ clathrate and water, and typical trajectories of a CH₄-rich fluid rising from a deep source up to the Martian surface and, ultimately, atmosphere. The top geothermal pressure-temperature profile (dashed blue line) uses a conductivity of 2.4 W m⁻¹ K⁻¹ and a density of 2 g cm⁻³, which is consistent with an ice-cemented soil. The bottom profile (solid blue line) uses a conductivity of 0.9 W m⁻¹ K⁻¹ and a density of 2.3 g cm⁻³, which is consistent with a dry sandstone. Properties are assumed constant with depth, with a surface temperature of 180 K and a geothermal flux of 30 mW m⁻² (from Mousis *et al*., 2013a).

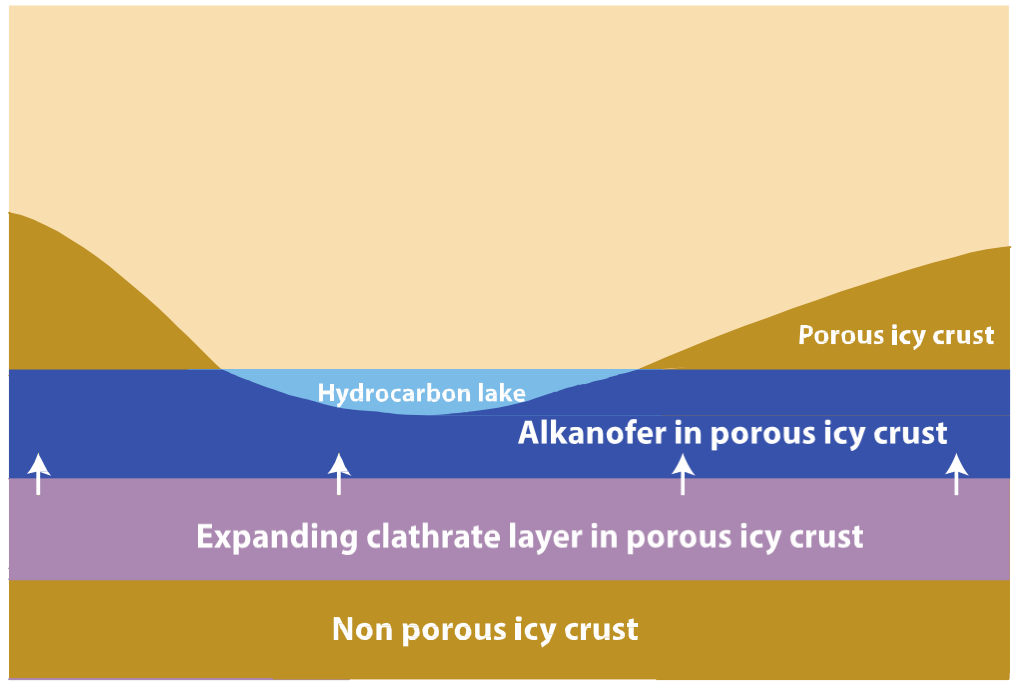

**Figure 7:** Schematic view representing the possible interaction between an alkanofer and a coexisting clathrate reservoir on Titan. The two phases are assumed to communicate and form an isolated system that occupies the open porosity of the icy crust. The clathrate reservoir expands with time throughout the porous icy crust at the expense of the  alkanofer.  The progressive transfer and fractionation of the molecules in the forming clathrate reservoir alter the  alkanofer's chemical composition (adapted from Mousis *et al*., 2014).

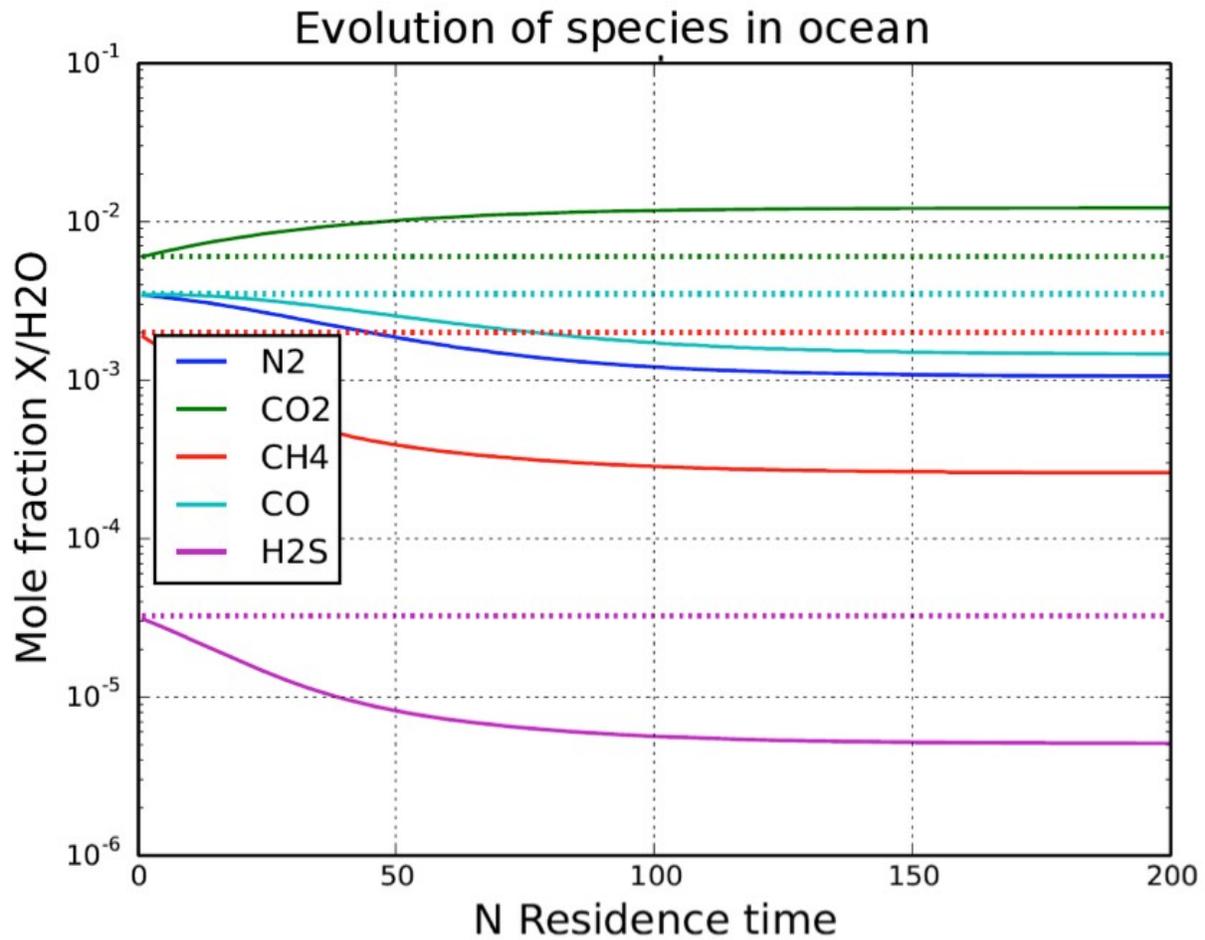

**Figure 8:** Evolution of the mole fraction of five species susceptible to encaging in the internal ocean of Enceladus. Dotted lines represent the plume mole fractions. Residence time refers to the time needed to renew the water of the ocean (model of subglacial lake - see Mousis *et al*., 2013c). This plot assumes that structure II clathrate is formed.